\documentclass[notitlepage,prd,twocolumn,showpacs,preprintnumbers,nofootinbib,superscriptaddress,floatfix,tightenlines]{revtex4-1}
\usepackage{mathrsfs}
\usepackage{amssymb}
\usepackage{amsmath}
\usepackage{graphicx}
\usepackage[utf8]{inputenc}
\usepackage{amsfonts,amsbsy}
\usepackage{nicefrac}
\usepackage{xcolor}
\definecolor{lcolor}{rgb}{0.5,0,0}
\definecolor{citcolor}{rgb}{0,0.3,0.0}
\usepackage[breaklinks,colorlinks,urlcolor=blue,citecolor=citcolor,linkcolor=lcolor]{hyperref}

\usepackage{comment}
\usepackage{bbm}

\usepackage[normalem]{ulem} 
\usepackage{enumitem}

\newcommand{\mbf}{\mathbf}

\newcommand{\mrm}{\mathrm}

\newcommand{\HTL}{\mrm{HTL}}
\newcommand{\fit}{\mrm{fit}}

\newcommand{\pInit}{Q}
\newcommand{\Q}{Q}
\newcommand{\wplas}{\omega_{\mrm{pl}}}
\newcommand{\gplas}{\gamma_{\mrm{pl}}}
\newcommand{\fig}{Fig.~}

\newcommand{\eq}{Eq.~}
\newcommand{\se}{Sec.~}
\newcommand{\eqs}{Eqs.~}
\newcommand{\re}{Ref.~}

\newcommand{\nr}[1]{(\ref{#1})}
\newcommand{\bs}[1]{\boldsymbol{#1}}

\newcommand{\ud}{\mathrm{d}}

\newcommand{\nn}{\nonumber}

\begin{document}

\title{Heavy quark diffusion in an overoccupied gluon plasma}

\author{K.~Boguslavski} 
\affiliation{Institute for Theoretical Physics, Technische Universität Wien, 1040 Vienna, Austria }

\author{A.~Kurkela} 
\affiliation{Theoretical Physics Department, CERN, Geneva, Switzerland}
\affiliation{Faculty of Science and Technology, University of Stavanger, 4036 Stavanger, Norway}

\author{T.~Lappi} 
\affiliation{Department of Physics, P.O.~Box 35, 40014 University of Jyv\"{a}skyl\"{a}, Finland}
\affiliation{Helsinki Institute of Physics, P.O.~Box 64, 00014 University of Helsinki, Finland}

\author{J.~Peuron} 
\affiliation{European Centre for Theoretical Studies in Nuclear Physics and Related Areas (ECT*) and
Fondazione Bruno Kessler, Strada delle Tabarelle 286, I-38123 Villazzano (TN), Italy}

\preprint{CERN-TH-2020-069}

\begin{abstract}
We extract the heavy-quark diffusion coefficient $\kappa$ and the resulting momentum broadening $\langle p^2 \rangle$ in a far-from-equilibrium non-Abelian plasma. We find several features in the time dependence of the momentum broadening: a short initial rapid growth of $\langle p^2 \rangle$, followed by linear growth with time due to Langevin-type dynamics and damped oscillations around this growth at the plasmon frequency. We show that these novel oscillations are not easily explained using perturbative techniques but result from an excess of gluons at low momenta. These oscillation are therefore a gauge invariant confirmation of the infrared enhancement we had previously observed in gauge-fixed correlation functions. We argue  that the kinetic theory description of such systems becomes less reliable in the presence of this IR enhancement.
\end{abstract}

\maketitle


\section{Introduction}

Transport coefficients, such as viscosities, diffusion coefficients and conductivities contain information about microscopic properties of the medium. In the framework of QCD matter produced in ultrarelativistic heavy-ion collisions, the evaluation of such transport coefficients has been a longstanding problem. Perturbative evaluations at Leading Order (LO) have been available for a long time  \cite{Arnold:2000dr,Arnold:2003zc,Hosoya:1983xm}. More recently perturbative calculations have been pushed to next-to-leading order (NLO) accuracy \cite{Ghiglieri:2018dib,Ghiglieri:2015zma,Ghiglieri:2018dgf,Ghiglieri:2013gia,Ghiglieri:2018ltw}. In equilibrium, there have been attempts  to extract transport coefficients also using nonperturbative lattice QCD methods \cite{Aarts:2014nba,Aarts:2007wj,Aarts:2002cc}.

Heavy quarks are unique probes of the transport properties of the quark gluon plasma (QGP) because of their large mass compared to the other scales of the medium. Pair production and annihilation processes are negligible, and all the heavy quarks within the medium are created in the hard processes preceding the formation of the QGP. 
Heavy quark observables carry information about the entire history of the medium.  

In conventional transport approaches to heavy-ion collisions, the effects of early-time, nonequilibrium evolution are usually ignored. Only very recently studies have addressed the importance of the nonequilibrium evolution. For heavy quark diffusion specifically, a Fokker-Planck approach  to the evolution of heavy quarks in a non-equilibrium gluon plasma or ``glasma'' present in the early stages of the evolution was used in~\cite{Carrington:2020sww,Mrowczynski:2017kso}. The authors find that the glasma phase can have a sizable contribution to momentum broadening and energy loss of heavy quarks. At later stages of the non-equilibrium evolution when the quasiparticle description is valid,  recent studies have indicated that the  pre-equilibrium effects can be important \cite{Das:2017dsh,Song:2019cqz}. In \cite{Ipp:2020mjc} jet momentum broadening in the glasma was investigated. The main result is that a colored particle can accumulate sizable momentum broadening during the glasma phase ($ \langle p_\perp^2 \rangle  = 1-4 \mathrm{GeV}^2$). One might thus expect the pre-equilibrium phase to be important also for heavy quarks.

The heavy quark momentum diffusion coefficient $\kappa$ can be studied in multiple ways. 
In thermal equilibrium, it has been calculated with perturbative methods~\cite{Arnold:2000dr,Arnold:2002zm,Arnold:2003zc,Moore:2004tg,CaronHuot:2007gq,CaronHuot:2008uh}
and studied with a standard lattice approach  \cite{Petreczky:2005nh,Meyer:2010tt,Francis:2011gc,Banerjee:2011ra,Francis:2015daa,Brambilla:2019oaa}. Another possibility is  to use lattice gauge theory in the classical approximation. This technique has been applied to the heavy quark diffusion coefficient $\kappa_\infty$ and the jet quenching coefficient $\hat{q}$ in thermal equilibrium systems~\cite{Laine:2009dd,Laine:2013lia,Panero:2013pla}. However, one of the benefits of the classical approach is that one can also study nonperturbative systems out of equilibrium, as we will do here.
Once the heavy quark diffusion coefficient is known, one can use it  to understand heavy quark flow and spectra by incorporating the diffusion  process in a simulation of the heavy ion collision \cite{Moore:2004tg,Akamatsu:2008ge,Rapp:2009my,vanHees:2005wb}.

The heavy quark diffusion coefficient $\kappa$ is not only important  for momentum broadening of heavy quarks, but it also has applications for quarkonia. Quarkonia can be modelled using an open quantum system approach \cite{Brambilla:2016wgg,Kajimoto:2017rel,Brambilla:2017zei,Akamatsu:2018xim,Brambilla:2019tpt}, and their time-evolution is governed by the Lindblad equation \cite{Lindblad:1975ef,Gorini:1975nb}. The equation of motion needs two transport coefficients as an input, one of which is the heavy-quark diffusion coefficient.

Our aim in this paper is to understand momentum broadening $\langle p^2 \rangle$ and the evolution of the momentum diffusion coefficient $\kappa$ of heavy quarks in a far-from-equilibrium overoccupied system, with the main motivation coming from initial stages in ultrarelativistic heavy ion collisions.
After the collision, occupation numbers of gluonic fields at the characteristic momentum scale $Q$ are non-perturbatively large $~\sim 1/g^2$ \cite{Lappi:2006fp,Gelis:2015gza} during initial stages in a weak-coupling thermalization picture. 
In this case, classical-statistical simulations are applicable and have been widely used 
\cite{Krasnitz:1998ns,Krasnitz:2000gz,Lappi:2003bi,Romatschke:2005pm,Romatschke:2006nk,Fukushima:2011nq,Kurkela:2012hp,Gelis:2013rba,Lappi:2016ato,Lappi:2017ckt,Mace:2016svc,Mace:2016shq,Berges:2017igc,Berges:2013eia,Berges:2013fga,Berges:2012ev,Schlichting:2012es,Gelfand:2016yho,Ipp:2017lho,Boguslavski:2019fsb,Ipp:2017uxo,Muller:2019bwd,Ipp:2018hai} to understand the pre-equilibrium dynamics in the collision. 

In this paper we simulate a highly occupied plasma in $SU(2)$ Yang-Mills theory in a three dimensional fixed box in a self-similar regime \cite{Kurkela:2011ti,Kurkela:2012hp,Berges:2013fga,Orioli:2015dxa,Berges:2014bba}. We extract the heavy-quark diffusion coefficient $\kappa$ and momentum broadening $\langle p^2 \rangle$ in this far-from-equilibrium system using suitable definitions of these gauge-invariant observables for out-of-equilibrium dynamics.  We explain how these generalized definitions can be used in quarkonium and diffusion studies and compare our results with perturbative calculations. 
Unlike e.g. in \cite{Ruggieri:2018rzi,Sun:2019fud,Liu:2019lac,Lehmann:2020kjg} we do not explicitly follow the motion of the quarks or quarkonia in the color field. Instead, we work in the infinite quark mass limit, where the quark is stationary, and measure the force acting on the quark from the chromoelectric field. 
More specifically, we will extract for the first time the heavy-quark diffusion coefficient in a classical non-equilibrium system. 
In addition to this, we find new features in the time-dependence of momentum broadening of a heavy quark.  While we consider a self-similar regime, the explanation of the different features in terms of a perturbative calculation is more general. Therefore similar features could be found in gluonic plasmas with other initial conditions, e.g., in the Glasma state in the initial stage of a heavy ion collision. In particular, we observe modulations of the growing $\langle p^2 \rangle$ with the plasmon frequency $\wplas$, which we are able to attribute to an excess of gluons at low momenta as compared to perturbative predictions. To our knowledge, this is the first time that such oscillations are connected to the gluonic IR enhancement that has been observed earlier in gauge fixed correlators~\cite{Boguslavski:2018beu}. Our result complements the observation of gauge-invariant condensation of \re\cite{Berges:2019oun} in the same non-Abelian systems as studied here.

This paper is structured as follows. First, in Sec.~\ref{sec:cascade} we will discuss the isotropic overoccupied gluonic system that we are studying. This system has been extensively analyzed in previous works, so we will be brief and concentrate on collecting the relevant numerical results and parametric time dependences that we will need in our subsequent analysis. We will then, in Sec.~\ref{sec:HQdiffCoef} discuss how the motion of heavy quarks in a dense gluonic system is related to the unequal time correlator of chromoelectric fields,
how it is connected to quarkonium and diffusion studies, 
and present our numerical calculation of this correlator in the overoccupied gluonic system. In Sec.~\ref{sec:timedep} we will construct two microscopic models for calculating these correlators from a momentum distribution of gluonic quasiparticles in the system, and compare our numerical results to these models in Sec.~\ref{sec:res}. We will then conclude with a brief discussion of our results in Sec.~\ref{sec:conc}.

\section{Highly occupied non-Abelian plasma}
  \label{sec:cascade}

The system we are studying is described by an $SU(2)$ pure gauge theory. Its classical evolution starts from an initial condition that is characterised by a single-particle occupation number distribution
\begin{align}
\label{eq_init_cond}
 f(t=0, p) = \frac{0.2}{g^2}\,\frac{\pInit}{p}\,e^{-\frac{p^2}{2\pInit^2}}.
\end{align}
The overall properties of this system are rather well understood from several studies~\cite{Berges:2008mr,Kurkela:2011ti,Kurkela:2012hp,Berges:2012ev,Schlichting:2012es,Berges:2013fga,York:2014wja}. We use the same initial conditions and numerical methods as in our previous paper~\cite{Boguslavski:2018beu}. Thus we will only briefly summarize the physical properties and the calculational methods of this system here, referring the reader to the references for more details.

The momentum scale $\Q$ controls the typical hard momentum in the initial distribution and we will write dimensionful quantities scaled with a suitable power of $\Q$. For low momenta $p \lesssim \Q$, occupation numbers are large $f(t,p) \gg 1$ and after a transient time, the system approaches a universal far-from-equilibrium attractor state, characterized by self-similar dynamics
\begin{align}
\label{eq_selfsim}
    f(t,p) = (\Q t)^{4\beta} f_s\left( (\Q t)^\beta p \right)\,.
\end{align}
The scaling exponent $\beta$ and scaling function $f_s(p)$ of this state are universal and insensitive to details of the initial conditions or to the precise value of the (weak) coupling. For the considered $d = 3$ spatial dimensions, one has $\beta = -1/7$. 

The momentum scale that dominates the energy density $\Lambda$, i.e., the momentum of hard excitations, grows with time as
\begin{align}\label{eq:lam}
    \Lambda \sim \Q (\Q t)^{-\beta} .
\end{align}
More precisely, we define the hard scale as the value of $p$ for which the integrand of the energy density $\propto p^2 \omega(p) f(t,p)$ is maximal, where the dispersion relation can be approximated by a relativistic dispersion $\omega(p) \approx \sqrt{p^2 + m^2}$. 
The (non-)thermal mass $m$ that gluons obtain in a medium can be computed perturbatively (and self-consistently) as 
\begin{align}
\label{eq_wplas_HTL}
    m^2 = 2 N_c \int \frac{\ud^3 p}{(2 \pi )^3}\, \frac{g^2 f(t,p)}{\omega(p)}\,,
\end{align}
where we have also included a mass correction in the denominator. The scale separation between this mass scale and the hard (temperature for a thermal system) forms the basis for the perturbative Hard-Loop (HTL) framework. The mass is connected to the Debye mass $m_D$ and to the plasmon frequency $\wplas$ via
\begin{align}
\label{eq:mD_from_wplas}
m_D^2 = 2 m^2\,, \quad \wplas^2 = \frac{2}{3}\, m^2.
\end{align}
The plasmon scale will turn out to be essential in our study of heavy quark diffusion far from equilibrium. Eq.~\eqref{eq_wplas_HTL} implies that $\wplas$ and $m_D$ decrease with time in the self-similar regime as
\begin{align}\label{eq:oplas}
    \wplas \sim m_D \sim \Q (\Q t)^{\beta} .
\end{align}
Physically, the plasmon frequency is the lowest energy that quasiparticle excitations can have and has been computed in \re\cite{Boguslavski:2018beu} as the peak position of the spectral function at vanishing momentum $\rho(\omega, p=0)$.

The quasi-particle peak of $\rho(\omega, p=0)$ is of Lorentzian form and has a width $\gplas \equiv \gamma(p=0)$, which corresponds to the damping rate of plasmon excitations. 
Its value for $\Q t = 1500$ and the same initial conditions as employed here has been extracted numerically from fits to $\rho(\omega, p=0)$ in \re\cite{Boguslavski:2018beu}. 
Using the HTL formalism, it can also be computed as $\gplas^\HTL \approx 6.64 N_c g^2 T_*/(24\pi)$ \cite{Braaten:1990it}. Here  $T_*$ is the effective temperature of the soft field modes, which is given by
\begin{equation}
\label{eq:effT}
g^2 T_*(t) = \frac{2 N_c}{m_D^2} \int \frac{\ud^3 p}{(2 \pi)^3} \left(g^2 f\right)^2(t,p)\,.
\end{equation}
Thus, in the scaling regime, we would expect the decay rate $\gplas$ to decrease with time like $g^2 T_*$, i.e., as 
\begin{align}  \label{eq:gplas}
    \gplas \sim g^2 T_* \sim \Q (\Q t)^{3\beta} .
\end{align}
From \eqs \nr{eq:lam}, \nr{eq:oplas} and \nr{eq:gplas} we obviously have, at late enough times, the hierarchy $\gplas \ll \wplas \ll \Lambda$, which is reminiscent of the hierarchy of scales $g^2 T \ll g T \ll T$ in thermal equilibrium with temperature $T$.
The extracted values in \re\cite{Boguslavski:2018beu} at $\Q t=1500$ are 
\begin{align}
\Lambda\left( \Q t =1500 \right) &= 2.1\, \Q \nonumber \\
m_D(\Q t = 1500) &= 0.21\,\Q \nonumber \\
g^2 T_*(\Q t = 1500) &= 0.03\,\Q \nonumber \\
\gplas(\Q t = 1500) &= 0.003\,\Q\,,
\label{eq_numbers_at_1500}
\end{align}
which indeed shows the expected separation of scales.

At momenta $p \lesssim \wplas$ we observed in our earlier work that the actual occupation number distribution~\cite{Boguslavski:2018beu} displays a feature that we refer to as an ``IR enhancement,'' a feature also seen in earlier studies (see, e.g., \cite{Berges:2012ev,Kurkela:2012hp}). By this term we mean that the occupation number is significantly larger than the behavior  $f(p) \sim T_*/p$ expected from perturbation theory. A gauge theory with a non-conserved number of particles is not expected to exhibit actual condensation (see, e.g.,~\cite{Kurkela:2012hp,York:2014wja,Blaizot:2016iir}), and we do not interpret this excess of gluons as an indication of condensation in the proper sense of the word. We will discuss this feature more quantitatively in Sec.~\ref{sec:equaltime}.

Due to the large occupation numbers $f \sim 1/g^2 \gg 1$, the non-perturbative quantum problem can be accurately mapped onto a classical-statistical lattice gauge theory with lattice spacing $a_s$ and lattice size $N_s^3$. Its far-from-equilibrium evolution can be then studied using computer simulations solving classical equations of motion in temporal axial gauge $A_0 = 0$. The equations are formulated in a gauge-invariant way using link fields $U_j(t,\mbf x) = \exp(ig\, a_s A_j(t,\mbf x)$, that replace the usual gauge fields $A_j(t,\mbf x)$ in the numerics, and chromo-electric fields $E_j(t,\mbf x)$. Since we use the same initial conditions and numerical method as in our previous paper \cite{Boguslavski:2018beu}, we refer the reader there and to references therein for details of our numerical approach. 

Data has been averaged over typically 10-15 configurations and error bars correspond to the standard error of the mean. 
If not stated otherwise, we use the lattice spacing $Q a_s=0.5$ and lattice sizes ranging from $N_s^3 = 128^3$ to $264^3$. We also vary $a_s$ and $N_s$ to check for possible lattice artifacts (see Appendix~\ref{sec_lattice_checks}).


\section{Heavy quark diffusion and momentum broadening}
\label{sec:HQdiffCoef}

\subsection{Heavy quark motion in a color field}
\label{sec_HQ_diffusion_intro}

We consider a heavy quark with mass $M$ in the highly occupied non-Abelian plasma far from equilibrium described above.  We take the quark mass to be the largest momentum scale in our system, i.e.,  larger than a typical hard momentum scale $M \gg \Lambda$. Consequently the formation time of the heavy quark is much shorter than any other timescale in the system. We can then assume that the production mechanism of the heavy quark factorizes from its interaction with the color field degrees of freedom. Thus we do not need to specify a particular production mechanism here, but just concentrate on the subsequent interactions of the quark with the medium. The large mass  of the quark also implies that the wave function is sufficiently localized so that we can use the classical equation of motion for its momentum:
\begin{align}
\label{eq_dp_evol}
 \dot{p}_i(t) = \mathcal{F}_i(t)\,.
\end{align}
Here $\dot{p}_i \equiv \frac{\ud p_i}{\ud t}$ and the force $\mathcal{F}_i$ is proportional to the chromo-electric field $E_i$ acting on the heavy quark. Averaging over color states of the quark and the ensemble of color field configurations gives a zero mean force acting on the quark $\langle\dot{p}\rangle = 0$. On the other hand, the variance of the force is given by the force-force correlator~\cite{Ipp:2020mjc}
\begin{align}
\label{eq_dpdp_heavyQuark}
 \langle \dot{p}_i(t)\dot{p}_i(t') \rangle &= g^2\,\frac{\text{Tr}\langle E_i(t) U_0(t,t') E_i(t') U_0(t',t) \rangle}{\text{Tr}\, \mathbbm{1}} \nonumber \\
 &= \frac{g^2}{2 N_c} \langle E_i^a(t) E_i^a(t') \rangle \\
 &\equiv \frac{g^2}{2 N_c} \langle E E \rangle(t,t') \,.
 \label{eq_EE_corr_def}
\end{align}
The electric field correlator is evaluated at the same spatial location because in the limit of large $M$, the velocity of the heavy quark is negligible. In the last line, we have defined the (statistical) correlation function $\langle E E \rangle(t,t')$. 
The trace is taken in the fundamental representation and in the last line we used temporal gauge with $U_0(t',t'') = \mathbbm{1}$ as well as a summation over repeated color indices of the adjoint representation $a = 1, \dots, N_c^2-1$. 

\begin{figure}
    \includegraphics[scale=0.5]{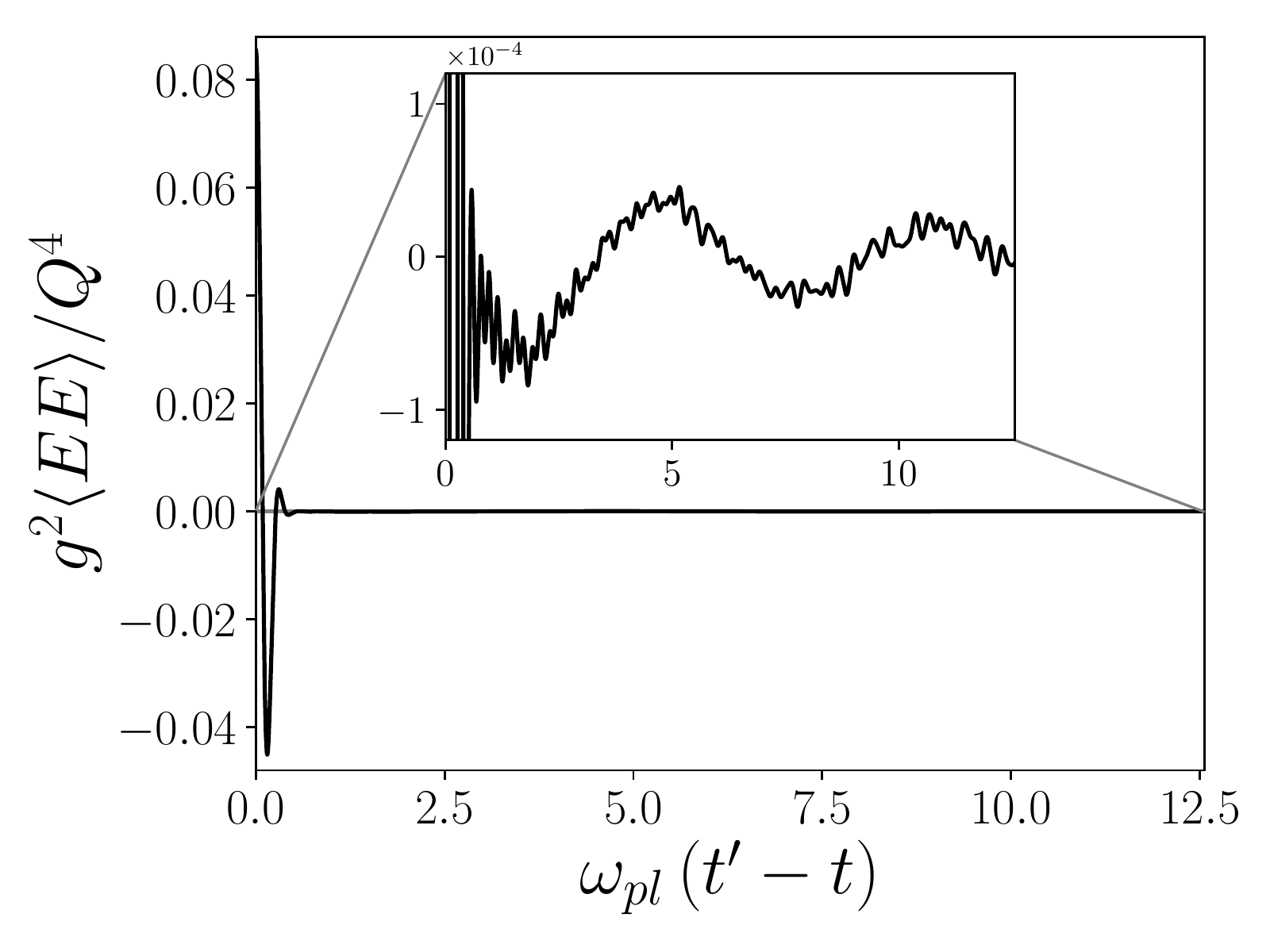}
    \includegraphics[scale=0.5]{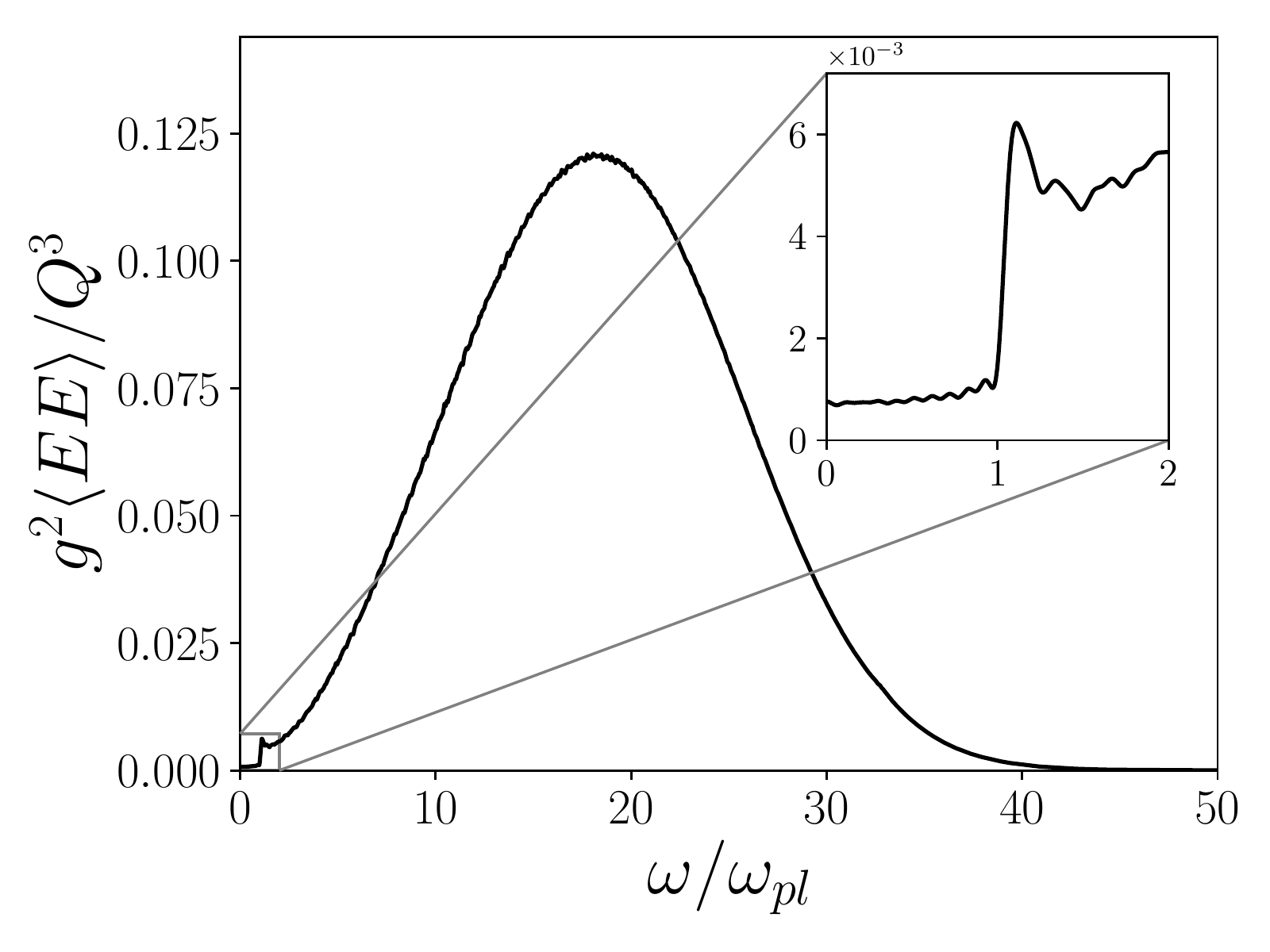}
    \caption{({\bf Left:}) The unequal time electric field correlation function \eqref{eq_EE_corr_def} as a function of relative time. The inset zooms into the data with a much smaller coordinate axis scale. One finds small oscillations with the plasmon frequency. ({\bf Right:}) The electric field correlator in Fourier space \eqref{eq_EE_corr_w}. The inset shows the low-frequency part of the curve. The visible structure can be understood using the gluon spectral function, as mentioned in Sec.~\ref{sec_HQ_diffusion_intro} and detailed in Sec.~\ref{sec:timedep}. 
    }
    \label{fig:EtE0signal}
\end{figure}

The correlation function \eqref{eq_dpdp_heavyQuark} is shown in \fig\ref{fig:EtE0signal} for the time $ \Q t = 1500$ as a function of the time difference $t' - t$. The signal starts with a large initial oscillation (upper panel) that quickly fades away on a time scale $\sim 1/\Lambda$. As shown in the inset, the correlator oscillates subsequently with a small frequency $\sim \wplas$. The amplitude of these oscillations is roughly a factor $10^3$ times smaller than the initial quick oscillations. 

It is useful to consider the Fourier transform of the correlator \eqref{eq_dpdp_heavyQuark} with respect to relative time $\Delta t = t'-t$. Normally we would define this in a symmetric way as
\begin{align}
\label{eq_EE_corr_w}
    &\langle E E \rangle(\bar{t},\omega) = \int_{-\infty}^{\infty} \ud \Delta t \; e^{-i\omega  \Delta t} \langle E E \rangle(\bar{t}-\Delta t/2,\bar{t}+\Delta t/2) \nonumber \\
    &\, = 2\,\text{Re} \int_{0}^{\infty} \ud \Delta t \; e^{-i\omega  \Delta t} \langle E E \rangle(\bar{t}-\Delta t/2,\bar{t}+\Delta t/2) 
\end{align}
for a fixed (central) time $\bar{t} \equiv (t+t')/2$, where we exploited that the statistical correlator is an even function. For the purpose of numerically extracting the frequency space correlation function for \fig\ref{fig:EtE0signal}, we approximate it by
\begin{align}
  \label{eq_EE_corr_w_approx}
  \langle E E \rangle(t,\omega) \approx 2\,\text{Re} \int_t^{t + \Delta t_{\text{max}}} \ud t'\, e^{-i\omega (t'-t)} \langle E E \rangle(t,t') 
\end{align}
for a fixed lower limit $t$. This is a good approximation as long as the time difference $|t-t'|$ is small compared to the rate of change as a function of the central time. In a power law cascade this latter can be estimated as the lifetime of the system, and thus the approximation is a good one when $t\approx t' \gg |t-t'|$, i.e., $\omega \gg 1/t$, which is the case here. 

The frequency space signal is shown in the lower panel of \fig\ref{fig:EtE0signal}. While it has a broad peak around $\omega \sim \Lambda$, the relevant part for heavy quark diffusion is located at low frequencies (see inset). This also illustrates a practical challenge related to the measurement: to obtain the low-frequency behavior correctly, high accuracy is required. The observed structure of the low-frequency part can be easily understood with the help of the spectral function $\rho(\omega,p)$, which will be discussed below in Sec.~\ref{sec:timedep}. Here we note that the finite piece at $\omega = 0$ stems from Landau damping of longitudinally polarized gluonic fields. On the other hand, the steep rise at $\omega \approx \wplas$ results from quasiparticle excitations, which can only contribute for frequencies $\omega \gtrsim \wplas$.

\begin{figure}
\centering
\includegraphics[scale=0.5]{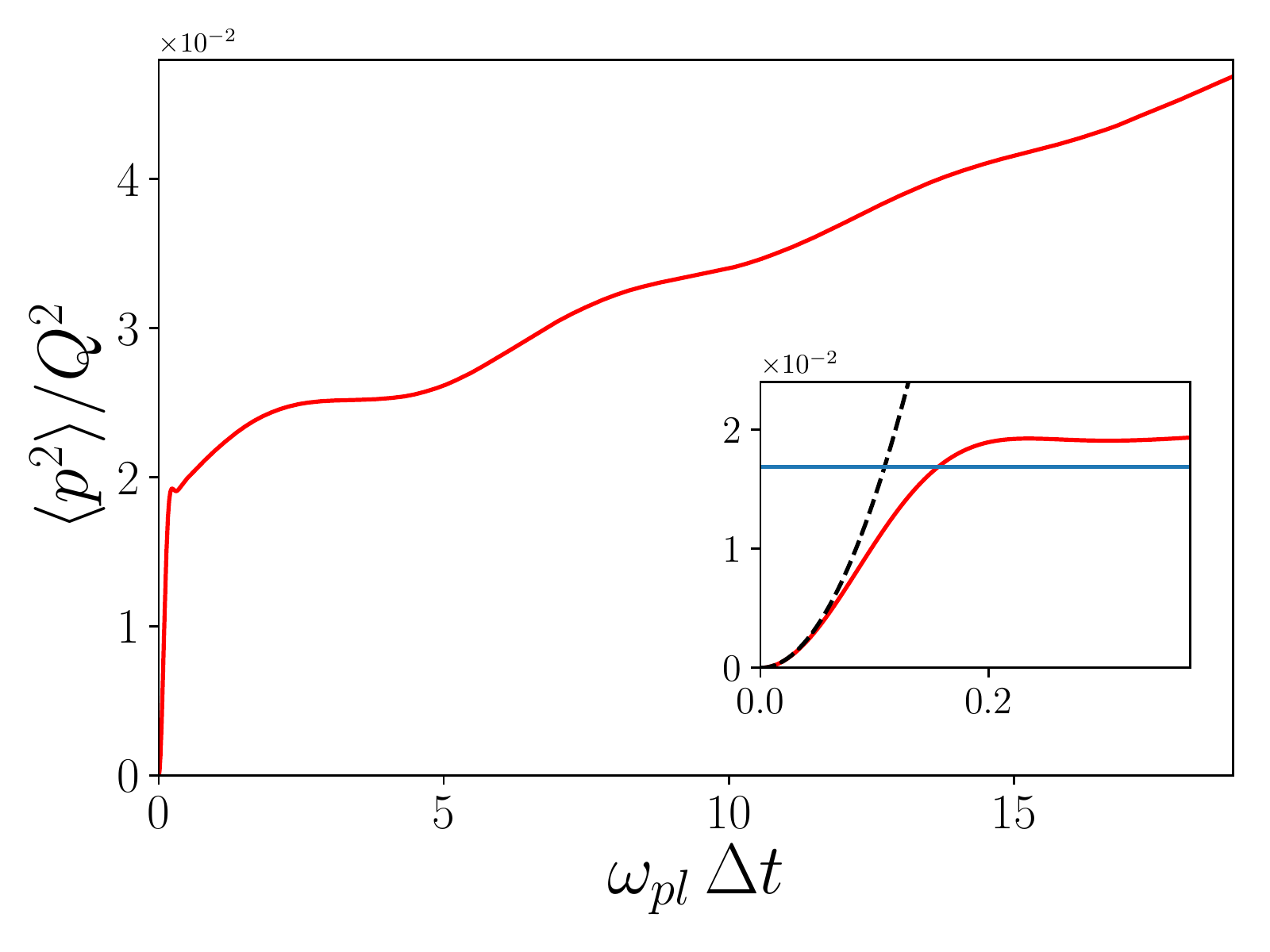}
\caption{The measured momentum broadening given by \eqref{eq_dpdp_heavyQuark}. We observe that after a rapid initial rise the momentum broadening increases roughly linearly in $p$. The linear rise is attributable to the interactions and the "steps" can be understood in the spectral reconstruction framework below. The initial rapid rise is shown in the inset with higher resolution. It corresponds to a decoherence effect whose early and later $\Delta t$ behavior of \eqs\eqref{eq:psqr_earlyDT} and \eqref{eq:psqr_lateDT} is shown as black-dashed and blue horizontal lines, respectively.
}
\label{fig:mombroadening}
\end{figure}

\subsection{Momentum broadening}
\label{sec_pSqr_broad}

So far, we have discussed the force-force correlation, which corresponds to $\langle \dot{p}_i(t')\dot{p}_i(t'') \rangle$ of a heavy quark traversing the non-equilibrium gluon plasma (see \eq\eqref{eq_dpdp_heavyQuark}). Integrating it, we arrive at the momentum broadening of the heavy quark after its creation at time $t$ as
\begin{align}
\label{eq_pp_heavyQuark}
 \langle p^2(t, \Delta t) \rangle = \frac{g^2}{2 N_c} \int_{t}^{t+\Delta t} \ud t' \int_{t}^{t+\Delta t} \ud t'' \,\langle E E \rangle(t',t'') \,.
\end{align}
This gauge-invariant physical observable is shown in \fig \ref{fig:mombroadening}. Its evolution shows three important features that are associated with different time scales for $\Delta t$: 
\begin{enumerate}[label=(\roman*)]
    \item \label{it:feat1} rapid growth at a short time scale of the order of the inverse hard scale $\Delta t \approx 2\pi/\Lambda$ ;
    \item \label{it:feat2} damped oscillations with period $\Delta t \approx 2\pi/\wplas$ ;
    \item \label{it:feat3} overall approximately linear growth $\,\sim \Delta t$ for $1/\Lambda \ll \Delta t \ll t$. 
\end{enumerate}
Each of these properties of momentum broadening has a different physical explanation and we will elaborate on them in this work. The modulations with frequency $\wplas$ in \ref{it:feat2} are a new feature that is, to our knowledge, observed in this work for the first time. We will discuss the relation of these oscillations to the quasiparticle properties of the plasma in \se\ref{sec:SRmethod} and show (see \fig\ref{fig:kappa_infty_t}) that they are related to an enhancement of infrared modes in the system over the perturbative expectation. 

While we will study these features \ref{it:feat1} - \ref{it:feat3} in detail below, we argue here that the different time scales result from the structure of $\langle E E \rangle(\bar{t},\omega)$. With the approximation that this correlator depends weakly on the central time $\bar{t} = (t'+t'')/2$ within the integration intervals, we can write
\begin{align}
    \langle p^2(t, \Delta t) \rangle 
     &  = \frac{g^2}{2 N_c} \int_{t}^{t+\Delta t} \!\!\!\! \ud t' \int_{t}^{t+\Delta t} \!\!\!\! \ud t'' \,\int_{-\infty}^{\infty} \frac{\ud \omega}{2\pi}\;  \nonumber \\ 
     & \times e^{i\omega (t'' - t')}\,\langle E E \rangle \left(\bar{t},\omega\right) \nonumber \\
    &  \approx \frac{g^2}{2 N_c} \int_{-\infty}^{\infty} \frac{\ud \omega}{2\pi}\; 4\,\frac{\sin^2(\omega\, \Delta t/2)}{\omega^2}\, \langle E E \rangle(t,\omega)\,.
    \label{eq_p2_wInt_EE}
\end{align}
With the frequency space correlator $\langle E E \rangle(t,\omega)$ shown in  \fig\ref{fig:EtE0signal}, we can distinguish the features based on different regimes of  $\Delta t$. When $\Delta t \lesssim 1/\Lambda$, basically all frequencies are included in the integral. Then the broad peak of $\langle E E \rangle(t,\omega)$ dominates the integration, which corresponds to frequencies with $\omega \sim \Lambda$ and the rapid growth observed under \ref{it:feat1}. For larger time scales $1/\Lambda \ll \Delta t \sim 1/\wplas$, the broad peak provides a constant shift in $\langle p^2 \rangle$ and the evolution of the integral is instead dominated by lower frequencies like those depicted in the inset of \fig\ref{fig:EtE0signal} (bottom). This corresponds to the properties  \ref{it:feat2} and  \ref{it:feat3}. 

We can compute the early rapid growth of \ref{it:feat1} analytically. For that, we remind ourselves that the fields in the correlator $\langle EE \rangle$ are evaluated at the same location $\mbf x = \mbf x'$. Due to spatial translation invariance, we can hence write\footnote{We use here the same approximation of replacing the central time with the initial one that was used in  \eq\nr{eq_p2_wInt_EE}.}
\begin{align}
    \langle p^2(t, \Delta t) \rangle \nonumber 
    & \approx \, \frac{g^2}{2 N_c}\int_{-\infty}^{\infty} \frac{\ud \omega}{2\pi}\; 4\,\frac{\sin^2(\omega\, \Delta t/2)}{\omega^2}\, \int \frac{\ud^3 p}{(2\pi)^3}\, \nonumber \\
    & \times \langle E E \rangle(t,\omega,p) \nonumber \\ 
    &\approx \,\frac{g^2}{2 N_c}\int \frac{\ud^3 p}{(2\pi)^3} \int_{-\infty}^{\infty} \frac{\ud \omega}{2\pi}\; 4\,\frac{\sin^2(\omega\, \Delta t/2)}{\omega^2} \, \nonumber \\
    & \times \langle E E \rangle(t,t,p) ~ \omega\,\rho(t,\omega,p) \nonumber \\
    &\approx \frac{4 g^2(N_c^2-1)}{N_c}\, \int \frac{\ud^3 p}{(2\pi)^3}\, \frac{f(t,p)}{\omega_{p}} \, \sin^2\left(\frac{\omega_{p} \Delta t}{2}\right) .
    \label{eq_psqr_init}
\end{align}
In the second line we used a generalized fluctuation-dissipation relation (given by \eqref{eq_fluct_diss_rel} in \se \ref{sec:SRmethod} where the fluctuation dissipation relation is discussed in more detail) to express the frequency space $\langle E E \rangle$ correlator in terms of the equal time $\langle E E \rangle$ correlator and the spectral function in frequency space. In our earlier paper~\cite{Boguslavski:2018beu} we have verified that such a relation holds very well in the system considered here. Since at small $\Delta t$ the integral is dominated by high frequencies, we then used the spectral function of free quasi-particles
\begin{align}
    \rho(t,\omega,p) \approx \rho^{\text{free}}(\omega,p) = 2\pi\, \text{sgn}(\omega) \, \delta(\omega^2-\omega_p^2)
\end{align}
and expressed the equal time correlator in terms of the distribution $f(t,p)$ as
\begin{align}
    \langle E E \rangle(t,t,p) = 2(N_c^2-1)\; \omega_p\,f(t,p).
    \label{eq_EE_fp_def}
\end{align}
Here $2(N_c^2-1)$ counts the degrees of freedom of transverse gluons and $\omega_p$ is their dispersion relation. 

The last line of \eq\eqref{eq_psqr_init} already explains our observations of the initial rise. The integral over momenta is dominated by the hard scale and hence, $\omega_{p} \sim p \sim \Lambda$. At early times $\Delta t \lesssim 1/\Lambda$, one can approximate 
\begin{align}
\label{eq:psqr_earlyDT}
    \langle p^2(t, \Delta t) \rangle \approx \frac{g^2}{2 N_c}\, (\Delta t)^2\, \varepsilon\,,
\end{align}
with the energy density $\varepsilon = 2(N_c^2-1)\int \ud^3 p \, \omega_{p} f(t,p) / (2\pi)^3$. This is shown as the black dashed curve in the inset in \fig \ref{fig:mombroadening}. At later times $\Delta t \gg 1/\Lambda$, the momentum integral involves rapid oscillations such that the approximation $\sin^2 (\omega_{p} \Delta t / 2) \approx 1/2$ can be used, leading to
\begin{align}
\label{eq:psqr_lateDT}
    \langle p^2(t, \Delta t) \rangle \approx \frac{3 (N_c^2-1)}{2 N_c^2}\, \wplas^2\,.
\end{align}
This is shown as the blue horizontal line in the inset in \fig \ref{fig:mombroadening}. Therefore, the initial fast growth stops due to decoherence after $\Delta t \sim 1/\Lambda$ at a value $\,\sim \wplas^2$.

\subsection{Heavy quark diffusion}
\label{sec_diff_kappa}

We can define the heavy-quark diffusion coefficient as the time derivative of the accumulated squared momentum \eqref{eq_pp_heavyQuark} at late times. For that purpose, let us define the function $\kappa\left(t , \Delta t \right)$ as follows
\begin{align}
\label{eq_def_kappa_der}
 3\kappa\left(t , \Delta t \right) &\equiv \frac{\ud}{\ud \Delta t}\langle p^2(t, \Delta t) \rangle  \\
 &= \frac{g^2}{N_c} \int_{t}^{t + \Delta t} \ud t' \langle E E \rangle (t+\Delta t,t') \nonumber \\
 &   \approx \frac{g^2}{N_c}\int_{-\infty}^{\infty} \frac{\ud \omega}{2\pi}\; \frac{\sin(\omega\, \Delta t)}{\omega}\,\langle E E \rangle(t,\omega),
 \label{eq_kappa_wInt_EE}
\end{align}
where we again approximated  the dependence on the central time in the correlator  $\langle E E \rangle(t,\omega)$. As we will see in the following, the limit $\Delta t\to \infty$ of $\kappa\left(t , \Delta t \right)$ gives the quantity that is commonly known as the \emph{heavy quark diffusion coefficient}.

The expression \eqref{eq_def_kappa_der} has also an interpretation for quarkonium evolution and decay in a non-Abelian plasma. As detailed in \re\cite{Brambilla:2016wgg}, it is related to the real part of the color-singlet self-energy (note that we have a different definition of the time arguments)
\begin{align}
\label{eq_ReSigma_kappa}
    \text{Re}\,\Sigma_s(t,\Delta t) &= \frac{g^2}{6 N_c}\,r^2 \int_{t}^{t + \Delta t} \ud t' \langle E_i^a(t+\Delta t) E_i^a(t') \rangle \nonumber \\
    &  = \frac{r^2}{2}\, \kappa\left(t , \Delta t \right),
\end{align}
where $r$ is the distance between the heavy quark and anti-quark. It is also proportional to the decay width $\Gamma$ in thermal equilibrium if $\Delta t$ is larger than any other time scale of the system. We emphasize, however, that the relation in the last line of \eqref{eq_ReSigma_kappa} is more general and also holds for finite $\Delta t$. 

It is beneficial to understand how the definition of $\kappa\left(t , \Delta t \right)$ in \eqref{eq_def_kappa_der} is related to the usual definition of the heavy-quark diffusion coefficient $\kappa^{\text{therm}}$ in thermal equilibrium (see, e.g., \cite{CaronHuot:2007gq}). Since thermal equilibrium is time translation invariant, there is no dependence on $t$ and we can simply set $t = 0$. Thermal equilibrium is also time reversal invariant and therefore the correlator $\langle E E \rangle$ is an even function of the time difference. Then we have
\begin{align}
\label{eq_kappa_th_def}
    \kappa_\infty^{\text{therm}} &\equiv \frac{g^2}{3 N_c}\,\int_{-\infty}^{\infty} \ud t'\, \langle \text{Tr}\,E_i(0) U_0(0,t') E_i(t') U_0(t',0) \rangle \nonumber \\
    &= \lim_{\Delta t \rightarrow \infty}\, \frac{g^2}{6 N_c}\,\int_{- \Delta t}^{\Delta t} \ud t'\, \langle E_i^a(0) E_i^a(t') \rangle \nonumber \\
     & = \lim_{\Delta t \rightarrow \infty}\, \frac{g^2}{3 N_c}\,\text{Re}\int_{0}^{\Delta t} \ud t'\, \langle E_i^a(0) E_i^a(t') \rangle \nonumber \\
&= \lim_{\Delta t \rightarrow \infty}\,\kappa\left(0, \Delta t \right),
\end{align}
where the superscript ``therm'' is a reminder of the thermal state considered here while the subscript $\infty$ corresponds to the limit $\Delta t \rightarrow \infty$. We note that in the literature, this coefficient is usually referred to as $\kappa$. In the second line of \eq\eqref{eq_kappa_th_def}, we used temporal gauge ($U_0 = \mathbbm{1}$). 
Obviously, we lose exact time translation invariance once we consider far-from-equilibrium systems as in the present work. However, we can still compute a time-dependent heavy-quark diffusion coefficient $\kappa_\infty(t)$ in analogy to \eq\eqref{eq_kappa_th_def} as
\begin{align}
\label{eq_kappa_infty_def}
    \kappa_\infty(t) = \left.\kappa\left(t , \Delta t \right)\,\right|_{t \gg \Delta t \gg 1/\gplas}\,.
\end{align}
Here the limit $\Delta t \rightarrow \infty$ is replaced by the condition that $\Delta t$ is larger than the longest life-time of quasi-particles in the plasma, which is given by the inverse damping rate of the zero mode $\gplas$. This is sufficient to assure that no contributions from quasi-particles enter the definition of $\kappa_\infty(t)$.  We cannot, however, formally take the infinite time difference limit, but  require $ \Delta t \ll t$. This is done so that the dependence of the correlator $\langle E E \rangle(t,t')$ on the central time is weak compared to its dependence on the time difference.
We note that, as for the thermal case \cite{CasalderreySolana:2006rq,CaronHuot:2009uh}, $3\kappa_\infty(t)$ is the zero-frequency value of the Fourier transform of the force-force correlator shown in \fig\ref{fig:EtE0signal}
\begin{align}
    \frac{g^2}{2 N_c} \langle E E \rangle (t, \omega = 0)~ =~ 3\,\kappa_\infty(t)\,.
\end{align}

The transport coefficient $\kappa_\infty(t)$ enables us to formulate a Langevin equation for heavy quarks in analogy to thermal equilibrium. Since the integral of the force-force correlator over a long but not infinite time interval $t \gg \Delta t \gg 1/\gplas$ is a constant, we can approximate the dynamics of the heavy quark at this timescale by a random momentum kick with the same normalization
\begin{align}
\label{eq_FF_Langevin}
 \langle \mathcal{F}_i(t') \mathcal{F}_j(t'') \rangle &= \frac{g^2}{2 N_c} \langle E_i^a(t) E_j^a(t') \rangle \nn \\ & \approx \kappa_\infty(t)\; \delta_{ij} \delta(t-t')\,.
\end{align}
This leads to a physical picture of the dynamics of a heavy quark in the medium as a Langevin process, which is commonly used in phenomenological applications.

Finally, we explain here how we extract $\kappa\left(t, \Delta t \right)$ from our simulations. In this paper we only consider sufficiently late times well within the self-similar regime so that we can stay in the limit $\Delta t \ll t$. Then one can, for the purpose of convenience in the numerical evaluation,  replace our definition \nr{eq_def_kappa_der} with a version where one of the electric fields is always evaluated at the lower,  instead of the upper, time 
\begin{align}
3\kappa\left(t , \Delta t \right)  &= 
\frac{g^2}{N_c} \int_{t}^{t + \Delta t} \ud t' \langle E E \rangle (t+\Delta t,t') \nonumber \\
 &   \approx 
\frac{g^2}{N_c} \int_{t}^{t + \Delta t} \ud t' \langle E E \rangle (t,t') .
 \end{align}
Also averaging over the (lattice) volume leads us to the equation
\begin{align}
\label{eq:kappa_from_data}
    \kappa\left(t, \Delta t \right) &\approx \frac{g^2}{3 N_c} \int_{t}^{t + \Delta t} \ud t' \int\frac{\ud^3 x}{V} \langle E_i^a(t,\mbf x) E_i^a(t',\mbf x) \rangle,
\end{align}
which we employ in our numerical extraction of $\kappa\left(t, \Delta t \right)$.

\begin{figure}
\centering
\includegraphics[scale=0.5]{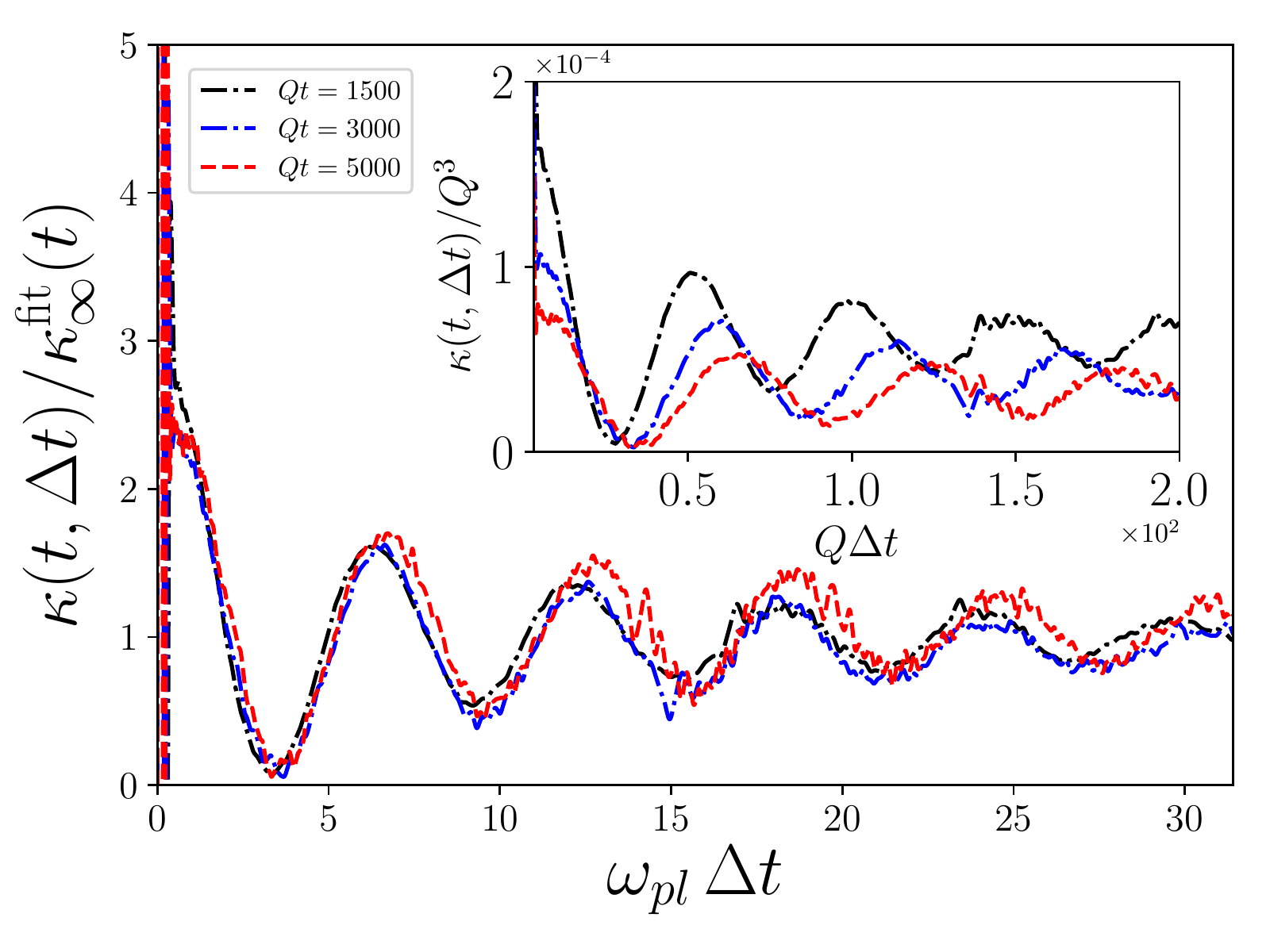}
\caption{Transient time behavior of the electric-field correlator $\kappa\left(t, \Delta t \right)$ as a function of $\Delta t$ extracted using the real time lattice method for various starting times. In the inset, the axes are scaled with the constant scale $Q$. In the main figure we rescale the vertical axis with the heavy-quark diffusion coefficient, which is the asymptotic $\Delta t\to \infty$ value $\kappa_{\infty}^{\mathrm{fit}}(t)$. It is extracted using the fit given by \eqref{eq_kappa_osc_fit} and its time dependence is shown below, in Fig.~\ref{fig:kappa_infty_t}. The horizontal axis is rescaled with the plasmon frequency $\wplas$ (effectively $t^{\nicefrac{-1}{7}}$) for each time $t$. That all curves fall on top of each other after rescaling is a sign of self-similarity in $t$.}
\label{fig:many_moneyplots}
\end{figure}

\begin{figure}
\centering
\includegraphics[scale=0.5]{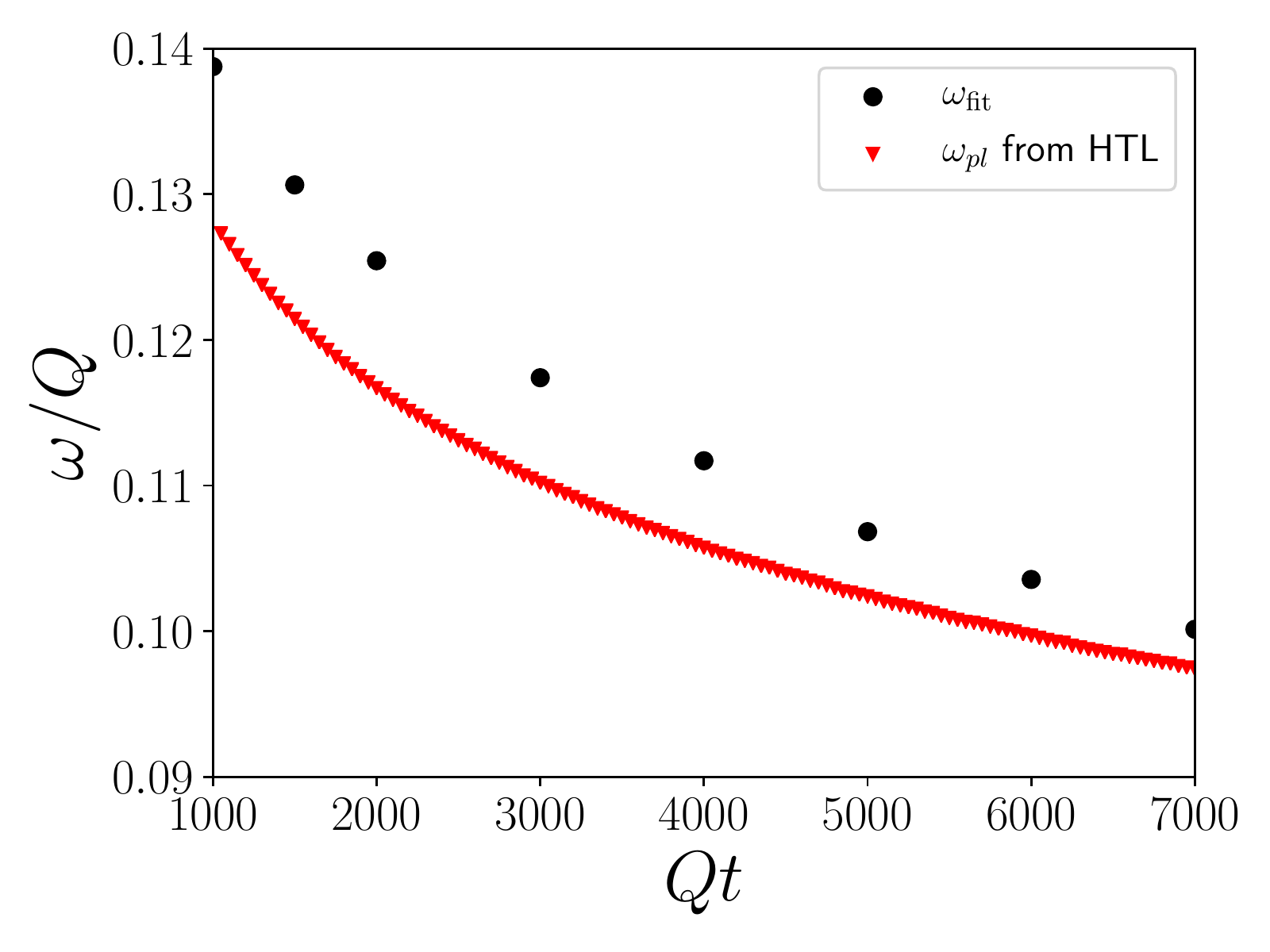}
\caption{Comparison of the plasmon mass scale $\wplas$ to the frequency $\omega_{\text{fit}}$ extracted from the unequal time electric field correlation function $\kappa\left(t, \Delta t \right)$ given by \eqref{eq:kappa_from_data}. The plasmon mass scale is extracted using \eq\eqref{eq_wplas_HTL}. The oscillation frequency of $\kappa\left(t, \Delta t \right)$ is extracted by fitting it to the damped oscillator in \eq\eqref{eq_kappa_osc_fit}. The main observation is that the frequencies are closely related.
}
\label{fig:osc_freq}
\end{figure}

\subsection{Self-similar behavior of $\kappa\left(t , \Delta t \right)$}
\label{sec_kappa_selfsim}

In \se\ref{sec_pSqr_broad} we have seen that after a quick initial growth, $\langle p^2(t, \Delta t) \rangle$ grows more slowly with time $\Delta t$, approximately linearly, and involves damped oscillations. We summarized these observations under the features of damped oscillations \ref{it:feat2} around a linear growth \ref{it:feat3}. Let us now study these properties in more detail in terms of $\kappa\left(t , \Delta t \right)$. We recall that this quantity can be thought of equivalently as  the time derivative of $\langle p^2(t, \Delta t) \rangle$ or as the integral of the electric field correlator over the time difference up to $ \Delta t$, and that its $\Delta t\to \infty$ limit is the heavy quark diffusion coefficient. 

The $\Delta t$-dependence of $\kappa\left(t , \Delta t \right)$  at  different times $\Q t = 1500$, $3000$, $5000$ is shown in \fig\ref{fig:many_moneyplots}.
The inset of \fig\ref{fig:many_moneyplots} shows our result scaled by the constant hard scale $\Q$ only.
 The main plot shows the correlator divided by the heavy-quark diffusion coefficient $\kappa_{\infty}^{\mathrm{fit}}(t)$, i.e. the $\Delta t\to \infty$ limit, which is extracted using the fit given by \eqref{eq_kappa_osc_fit}. As we will show in Sec.~\ref{sec:res}, the time dependence of $\kappa_{\infty}(t)$ can be well described by a functional form of  
$(\Q t)^{-5/7}$ times a logarithm of time that we will motivate in \se\ref{sec:timedep} (see the explicit expression given in \eq\nr{eq:ourkappa}).
The correlators in the main plot are plotted as a function of the time difference scaled by the plasmon frequency, $\wplas(t) \Delta t$. 
The plasmon frequency used for the rescaling is computed using the HTL formulas \eqref{eq_wplas_HTL} and \eqref{eq:mD_from_wplas}, which amounts to effectively rescaling the horizontal axis with a power law $(\Q t)^{-1/7}$. As explained in Sec.~\ref{sec:cascade}, the different values of $t$ correspond to different ratios of the physical scales in the problem. 
The fact that the curves from different times $t$ overlap as functions of rescaled time, clearly shows that the oscillations in $\kappa\left(t , \Delta t \right)$ happen at a scale determined by the plasmon frequency.
The scaling of the amplitude of the oscillations in \fig\ref{fig:many_moneyplots} shows that this same time dependence also describes the amplitudes of the oscillations as a function of $\Delta t$, a sign of self similar evolution in $\kappa\left(t , \Delta t \right)$.

The oscillatory form in  \fig\ref{fig:many_moneyplots} can be fitted for $\Delta t \gtrsim 1/\Lambda$ separately for each $t$ to a damped harmonic oscillator with a constant offset term
\begin{align}
\label{eq_kappa_osc_fit}
    \kappa_\fit\left(t , \Delta t \right) \approx \kappa_\infty^\fit(t) + A_\fit \cos(\omega_\fit \Delta t - \phi_\fit)\, e^{-\gamma_\fit \Delta t}.
\end{align}
From this fitting procedure, we extract the frequency $\omega_\fit(t)$ and show it in \fig\ref{fig:osc_freq} together with the computed frequency $\wplas^\HTL$ as functions of time. Since the frequencies are quantitatively close to each other, we conclude 
\begin{align}
    \omega_\fit \approx \wplas(t)\,.
\end{align}
Also the heavy-quark diffusion coefficient $\kappa_\infty(t)$ is extracted using the fit in \eqref{eq_kappa_osc_fit}. According to \eq\eqref{eq_kappa_infty_def}, the coefficient $\kappa_\infty(t)$ is defined as the late relative time limit $\Delta t \gg 1/\gplas$ of $\kappa\left(t, \Delta t \right)$. For finite $\Delta t$, it corresponds to the offset $\kappa_\infty^\fit(t)$ of the oscillations visible in \fig\ref{fig:many_moneyplots} and incorporated into the fit function \eqref{eq_kappa_osc_fit}. Measuring $\kappa_\infty(t)$ as $\kappa_\infty^\fit(t)$ reduces the residual dependence on $\Delta t$ and will be used as our standard way to extract $\kappa_\infty(t)$. 

We have also studied lattice regularization effects on our extraction of $\kappa_\infty(t)$, with results shown in Appendix~\ref{sec_lattice_checks}. We find that  $\kappa$ is insensitive to the IR cutoff. In the case of the UV cutoff, the results start to drift considerably for lattice spacings larger than $Q a_s > 0.6$ and hence, we use smaller lattice spacings. 
The Debye scale stays well within the reach of our lattice at all times.

\section{Understanding the time dependence of the correlator}
\label{sec:timedep}

To understand the observations in the previous section in terms of microscopic degrees of freedom in the system, we construct here two models for the $\Delta t$-dependence of $\kappa\left(t , \Delta t \right)$ and compare them to  our numerical results.  A crucial role in our discussion here is played by the momentum space gluonic equal-time correlation function $\langle E E \rangle(t,t,p)$ in Coulomb gauge, which we interpret in terms of a single particle distribution of gluons.  The input in the models that we want to construct is this single particle distribution that we extract from our simulation. This is an equal time correlator, meaning that it represents an integral over frequencies. It is also needed as a function of momentum, which means that in coordinate space it is not local. This single particle distribution is also not manifestly gauge invariant, but evaluated using field configurations in the Coulomb gauge. From this information we want to construct the heavy quark diffusion coefficient, which is local in coordinate space, i.e., involves an integral over gluonic momenta. The heavy quark diffusion coefficient is the zero frequency limit of an electric field correlator, i.e., requires correlators at unequal times.

We will here use two different approaches to go from equal time -- unequal coordinate correlations to an unequal time -- equal coordinate one.  The basic idea of the first one, that we call here the spectral reconstruction (SR) method, is to assume that the spectral functions in  the general momentum-frequency space are the ones given by (HTL) perturbation theory, and that they are related to the statistical function by a generalized fluctuation-dissipation relation.
This connection enables us to relate the statistical functions in different parts of phase space to each other, via the intermediary of the spectral function. The underlying assumptions are backed up by our previous numerical results in \re\cite{Boguslavski:2018beu}. 
The second one, of course related to the first one in the appropriate parametric regime, is to use known perturbative calculations of the heavy quark diffusion coefficient in kinetic theory, and simply substitute our measured single gluon distribution in such a calculation. We will first construct these two models in this section, and then compare them to the numerical result for the electric field correlator $\kappa\left(t , \Delta t \right)$ in \se\ref{sec:res}.

\begin{figure}
\centering
\includegraphics[scale=0.5]{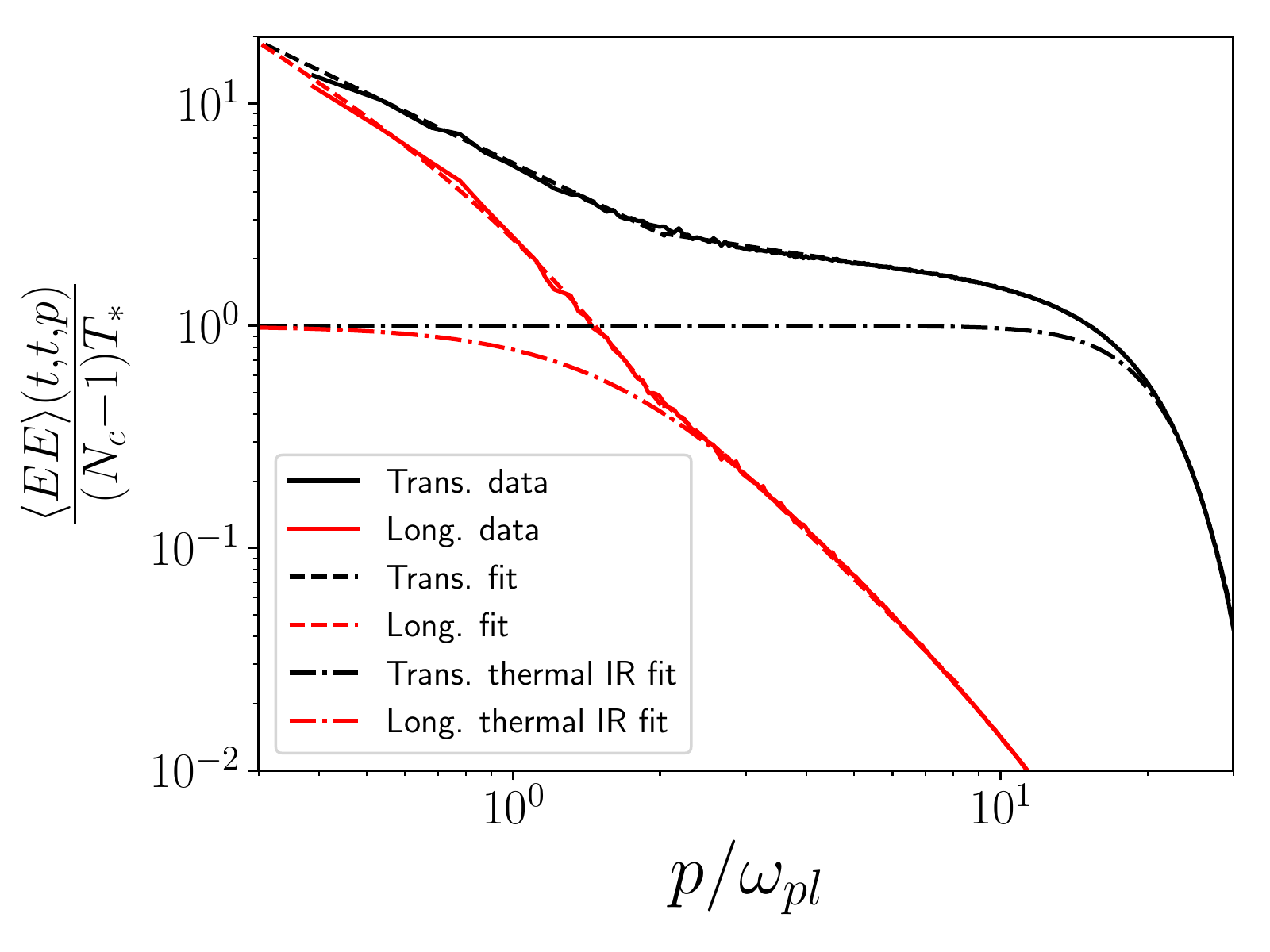}
\caption{Extracted equal time transverse and longitudinal statistical functions (continuous lines). The fits to the data are shown using dashed lines, corresponding to the infrared enhanced equal time statistical correlation function. The dash dotted lines correspond to the unenhanced correlator, which corresponds to the thermal IR expectation in the infrared given by \eqref{eq:thermalIR_definition}. The expectation is then smoothly matched to data to incorporate a proper UV behavior. }
\label{fig:statfun}
\end{figure}

\subsection{Equal time electric field correlator}
\label{sec:equaltime}

We start here with the statistical correlation function. 
In general it is defined as the anticommutator of Heisenberg field operators 
\begin{align}
 \label{eq_stat_fct}
 \langle E E \rangle_{jk} (x, x') = \frac{1}{2}\left\langle \left\{ \hat{E}_j^a(x), \hat{E}_k^a(x') \right\} \right\rangle,
\end{align}
with $x \equiv (t,\mbf x)$. 
In the classical-statistical approximation it becomes the expectation value of the  product 
\begin{align}
 \frac{1}{2}\left\langle \left\{ \hat{E}_j^a(x), \hat{E}_k^a(x') \right\} \right\rangle \rightarrow \left\langle E_j^a(x), E_k^a(x') \right\rangle\,.
\end{align}
On a periodic lattice the Fourier transform with respect to the relative coordinate $\mbf x-\mbf x'$ averaged over the whole lattice can conveniently be computed by Fourier transforming the electric fields  as $\langle E E \rangle(t,t',p) = \left\langle E_j^a(t,\mbf p), \left(E_k^a(t', \mbf p)\right)^* \right\rangle/V$. We refer to its Fourier transform to frequency space as $\langle E E \rangle_{jk}(t,\omega,p)$, neglecting the difference between fixed $t$ and fixed $\bar{t}=(t+t')/2$ as discussed in Sec.~\ref{sec:HQdiffCoef}.

It is necessary to distinguish transverse and longitudinal projections, which are defined as $2 P^T_{jk} = \delta_{jk} - p_j p_k/p^2$ and $P^L_{jk} = p_j p_k/p^2$, respectively, where $2$ counts the number of transverse polarizations. Correlators can then be decomposed into polarizations
\begin{align}
    \langle E E \rangle \equiv \langle E E \rangle_{jj} = 2\langle E E \rangle_{T} + \langle E E \rangle_{L}\,. 
    \label{eq_EE_polar}
\end{align}
In our previous publication \cite{Boguslavski:2018beu} we observed that the {\em{equal-time}} statistical correlation function $\langle E E \rangle_{T,L}(t,t,p)$ is enhanced compared to HTL expectations at low momenta for both polarizations. We show the numerically extracted $\langle E E \rangle_{T,L}(t,t,p)$ correlators in \fig\ref{fig:statfun} in the self-similar regime at time $\Q t = 1500$ as solid curves. The figure also shows in dashed lines a fit to these numerical results. To enable a more efficient evaluation of some integrals appearing below in  our models for the time-dependent correlator, we will in practice use these fits instead of our original numerical data. Moreover, using $\langle E E \rangle_T(t,t,p)$, we can define the distribution function as in our previous publication \cite{Boguslavski:2018beu} as
\begin{align}
\label{eq_def_fp}
    f(t,p) = \frac{1}{N_c^2-1}\,\frac{\langle E E \rangle_T(t,t,p)}{\sqrt{p^2 + m^2}}\,.
\end{align}
This is the definition that we used to extract the values~\nr{eq_numbers_at_1500} in Sec.~\ref{sec:cascade}, with an iterative procedure to simultaneously extract both $f(t,p)$ and $m$ using \eqs\nr{eq_wplas_HTL} and~\nr{eq_def_fp}. 

In HTL theory at leading order, the electric field correlator would, for low momenta $p \ll m_D$, be expected to approach  a constant 
\begin{align}
\label{eq:thermalIR_definition}
 \langle E E \rangle_{T,L}(t, t , p) \approx T_*\,\dot{\rho}_{T,L}(t, t , p) \approx T_* \, .
\end{align}
Note that the second ``$\approx$'' is in fact an equality at all $p$ for the transverse polarization but only at $p=0$ for the longitudinal one: for a discussion of the spectral function see Appendix~\ref{app:HTL}.
The value of this constant, i.e., the effective temperature of the infrared modes\footnote{One way to see that one expects a constant is to note that the equipartition of energy in thermal equilibrium  at temperature $T_*$ for a classical noninteracting theory would correspond to an expectation value $T_*/2$ for every quadratic term in the Hamiltonian, which in this case are (half) the squares of the components $E_j^a(t,\mbf p)$.} at low momenta, is conventionally denoted by $T_*$. 
We construct a  parametrization of this expected behavior by taking the parametrization of our data  at high momenta and smoothly matching it to a constant value determined by the temperature $T_*$ calculated using \eq\nr{eq:effT}. This ``thermal IR'' parametrization is shown by the dot-dash curves in \fig\ref{fig:statfun}. We emphasize that this second parametrization is meant to represent a scenario without the ``infrared enhancement'' seen in the correlator and discussed above, but keeping the large momentum degrees of freedom as close to the ones present in the lattice calculation as possible. We can then use both, the parametrization of our data including the infrared enhancement, and the one where it has been removed, to construct a model for the time-dependent correlators both with and without the infrared enhancement. Using this approach we will in fact argue below that the excess of gluons at low momenta is the main reason for the oscillations in $\Delta t$ observed in $\langle p^2 \rangle$ and $\kappa$.

\subsection{Spectral Reconstruction (SR) method}
\label{sec:SRmethod}

The aim of this section is to develop a spectral reconstruction (SR) method, which is based on our previous measurements of the gluonic spectral functions, the $\langle E E \rangle_{T,L}(t,t,p)$ \emph{equal-time} correlators and expectations from perturbation theory, and that  can be compared to our measurement of the \emph{unequal} time electric field correlators, specifically to our numerical result for $\kappa(t,\Delta t)$.

For this, let us reformulate $\kappa\left(t , \Delta t \right)$ using the full correlation functions $\dot{\rho}(t,\omega,p)$ and $\langle E E \rangle(t,\omega,p)$, respectively. We discussed the statistical correlation function in the previous subsection. The spectral function (strictly speaking the time derivative of the spectral function, but we will call $\dot{\rho}$ the spectral function in an abuse of language below) is defined as the commutator 
\begin{align}
 \label{eq_spectral_fct}
 \dot{\rho}_{jk} (x, x') = \frac{1}{N_c^2-1} \left\langle \left[ \hat{E}_j^a(x), \hat{A}_k^a(x') \right] \right\rangle. 
\end{align}
In the classical-statistical approximation, it can be computed using the retarded propagator $\dot{G}_{jk}^R(x,x') = \theta(t-t')\,\dot{\rho}_{jk} (x, x')$, which can be extracted from simulations with linear response theory \cite{Boguslavski:2018beu}. We denote the Fourier transform of $\dot{\rho}$ with respect to relative time and spatial coordinates as $\dot{\rho}_{jk}(t,\omega,p)$. Transverse and longitudinal polarizations can be distinguished as for the statistical correlator in \eq\eqref{eq_EE_polar}. For thermal equilibrium the fluctuation-dissipation relation states that the $\omega$-dependence of these functions is the same, i.e., the ratio $\langle E E \rangle_{T,L}(t, \omega , p)/\dot{\rho}_{T,L}(t, \omega , p)$ is a known function that only depends on the momentum  $p$ and
the temperature. We can easily generalize this to a nonequilibrium situation by taking this function of $p$ to be one determined by the equal-time statistical function (and the equal time spectral functions $\dot{\rho}_{T,L}(t, t , p)$, whose expressions are written in Appendix~\ref{app:HTL}).
Explicitly, we start by assuming that 
\begin{align}
 \label{eq_fluct_diss_rel}
 \frac{\langle E E \rangle_{T,L}(t, \omega , p)}{\langle E E \rangle_{T,L}(t, t , p)} = \frac{\dot{\rho}_{T,L}(t, \omega , p)}{\dot{\rho}_{T,L}(t, t , p)}
\end{align}
even out of equilibrium. We have indeed observed numerically in \re\cite{Boguslavski:2018beu} that this relation holds in the self-similar regime.

With these considerations, we can write \eq\eqref{eq_kappa_wInt_EE} as
\begin{align}
\label{eq_kappa_wInt_EE_pInt}
 & \!\!\!\!\!\!\!\! 3\kappa\left(t , \Delta t \right)  \\
 = &  \frac{g^2}{N_c}\int_{-\infty}^{\infty} \frac{\ud \omega}{2\pi}\; \frac{\sin(\omega\, \Delta t)}{\omega}\, \int \frac{\ud^3 p}{(2\pi)^3}\,\langle E E \rangle(t,\omega,p) \nonumber \\
 = \;& \frac{g^2}{N_c}\int \frac{\ud^3 p}{(2\pi)^3} \int_{-\infty}^{\infty} \frac{\ud \omega}{2\pi}\; \frac{\sin(\omega\, \Delta t)}{\omega}\, \nonumber \\
 & \times  ~\left[ 2\langle E E \rangle_{T}(t,\omega,p) + \langle E E \rangle_{L}(t,\omega,p) \right] \nonumber \\
 = \;& \frac{g^2}{N_c} \int \frac{\ud^3 p}{(2\pi)^3} \int_{-\infty}^{\infty} \frac{\ud \omega}{2\pi}\; \frac{\sin(\omega\, \Delta t)}{\omega}\, \nonumber \\
 & \times \left[ 2\langle E E \rangle_{T}(t,t,p)\frac{\dot{\rho}_{T}(t,\omega,p)}{\dot{\rho}_{T}(t,t,p)} + \langle E E \rangle_{L}(t,t,p)\frac{\dot{\rho}_{L}(t,\omega,p)}{\dot{\rho}_{L}(t,t,p)} \right].
 \nonumber
\end{align}
We now have to determine the spectral functions. Note that they appear as ratios, which are normalized to unity 
\begin{align}
    \int_{-\infty}^{\infty} \frac{\ud \omega}{2\pi}\;\frac{\dot{\rho}_{T,L}(t,\omega,p)}{\dot{\rho}_{T,L}(t,t,p)} = 1\,.
\end{align}
According to HTL calculations at LO, each spectral function can be decomposed in frequency space into  parts that are associated with Landau damping and to quasiparticle excitations, resulting in 
\begin{align}
    \dot{\rho}_{T,L}(t,\omega,p) = \dot{\rho}_{T,L}^{\mrm{Landau}}(t,\omega,p) + \dot{\rho}_{T,L}^{\mrm{QP}}(t,\omega,p)\,.
    \label{eq_rho_contributions}
\end{align}
We have seen this same structure in our numerical calculations of the spectral functions in  \re\cite{Boguslavski:2018beu}. 
The quasiparticle contributions can be written as
\begin{align}
\label{eq:rhopart}
    \dot{\rho}_{T,L}^{\mrm{QP}}(t,\omega,p) &= 2\pi\,Z_{T,L}(p)\;\omega\, \big[h_{T,L}\left(\omega - \omega_{T,L}(p), p\right) \nonumber \\
    & - h_{T,L}\left(\omega + \omega_{T,L}(p), p\right) \big],
\end{align}
in terms of the dispersion relations $\omega_{T,L}(p)$ of transversely and longitudinally polarized quasi-particles, the residues $Z_{T,L}(p)$ and the functions $h_{T,L}\left(\omega, p\right)$ that are normalized to unity and correspond to the quasiparticle peaks. Perturbatively, the damping rate of the quasiparticles is of the order $g^2 T_*$. 
Thus, at leading order HTL the quasiparticle peaks correspond to delta functions $h_{T,L}\left(\omega, p\right) \rightarrow \delta(\omega)$. More realistically, the quasiparticle peaks exhibit a Lorentzian shape with finite damping rate $\gamma_{T,L}(p)$
\begin{align} \label{eq:peak}
    h_{T,L}\left(\omega, p\right) = \frac{1}{\pi} \frac{\gamma_{T,L}(p)}{\omega^2 + \gamma_{T,L}^2(p)}\,,
\end{align}
which we have also verified numerically in \cite{Boguslavski:2018beu}. 
The explicit leading order expressions for dispersion relations $\omega_{T,L}(p)$, quasiparticle residues $Z_{T,L}(p)$ and Landau damping contributions $\dot{\rho}_{T,L}^{\mrm{Landau}}(t,\omega,p)/\dot{\rho}_{T,L}(t,t,p)$ are written in the Appendix~\ref{app:HTL}.

Combining \eqs\nr{eq_kappa_wInt_EE_pInt} and~\eqref{eq_rho_contributions}, we can split the diffusion coefficient \eqref{eq_kappa_wInt_EE_pInt} into four parts, corresponding to transverse / longitudinal and to Landau / quasiparticle contributions. Note that for each of these contributions, the frequency integration simplifies. The Landau damping contributions only have support for $|\omega| < p$, i.e.,  $\dot{\rho}_{T,L}^{\mrm{Landau}}(t,\omega,p) \propto \theta(p^2 - \omega^2)$. Thus the frequency integration becomes $\int_{-\infty}^{\infty} \ud \omega \mapsto \int_{-p}^{p} \ud \omega$ for these contributions. For the quasiparticle contributions the frequency integration can be done analytically as
\begin{align}
\label{eq:omegaint}
    &\int_{-\infty}^{\infty} \frac{\ud \omega}{2\pi}\, \frac{\sin(\omega\, \Delta t)}{\omega}\;2\pi\,Z_{T,L}(p)\;\omega\, \nonumber \\ 
    &\times \left[h_{T,L}\left(\omega - \omega_{T,L}(p), p\right) - h_{T,L}\left(\omega + \omega_{T,L}(p), p\right) \right] \nn \\
    &=\;  2\,Z_{T,L}(p)\,\sin(\omega_{T,L}(p)\, \Delta t)\,e^{-\gamma_{T,L}(p)\,|\Delta t|}\,.
\end{align}
All the remaining integrals are in general performed numerically. We have observed deviations from the HTL expressions at LO for $\omega_{T,L}(p)$, $\gamma_{T,L}(p)$ and $\langle E E \rangle_{T,L}(t,t,p)$. Thus, in our numerical calculations of $\kappa^{\text{SR}}(t, \Delta t)$ we will use the following forms, which we have extracted in our linear response framework \cite{Kurkela:2016mhu,Boguslavski:2018beu}:
\begin{itemize}
\item Since we observed for the dispersion relations $\omega_{T,L}(p)$ some deviations from the respective HTL expressions at LO, we use fits to our data from \cite{Boguslavski:2018beu}. 
\item We use the damping rates $\gamma_{T,L}(p)$  extracted in \re\cite{Boguslavski:2018beu}. However, as will be shown in \fig\ref{fig:HTL_break_down}, setting $\gamma_{T,L}(p) = 0$ does not change the results considerably.
\item For the statistical $\langle E E \rangle_{T,L}(t,t,p)$ correlation function we observe  significant enhancement over the HTL expectation in the infrared. To study this effect we use both, a parametrization of our numerical result for $\langle E E \rangle_{T,L}(t,t,p)$ and one where this enhancement has been removed, as discussed above in Sec.~\ref{sec:equaltime} and shown in \fig\ref{fig:statfun}.
\end{itemize}
For the other ingredients needed in the calculation: for the functional form of the quasiparticle peak $h_{T,L}$ we use the expression~\nr{eq:peak}, and for the Landau damping contribution $\dot{\rho}_{T,L}^{\mrm{Landau}}(t,\omega,p)$ and the quasiparticle residue $Z_{T,L}(p)$ we use the standard forms from the literature that can be found explicitly in Appendix~\ref{app:HTL}, with the value of the Debye mass obtained using the iterative procedure discussed above. Also note that in all of these expressions, a time dependence enters both due to $\langle E E \rangle_{T,L}(t,t,p)$ and due to the time dependence of the Debye mass $m_D(t)$.

We will evaluate the full $\Delta t$ dependence numerically with the procedure described above. It is also important to compute $\kappa_\infty(t)$, which emerges in the limit $\Delta t \to \infty$, i.e., the actual heavy quark diffusion coefficient. 
In this limit the sinc functions in the frequency integral \nr{eq_kappa_wInt_EE_pInt} become delta functions
\begin{equation} \label{eq:sincdelta}
\lim_{\Delta t \to \infty} \frac{2 \sin\left( \omega \Delta t \right)}{\omega} \rightarrow 2 \pi \delta(\omega).
\end{equation}
The transverse Landau damping contribution vanishes in the limit $\omega/p \to 0$ (see \eq\nr{eq:trlandau}  in Appendix~\ref{app:HTL})  and thus does not contribute. The longitudinal Landau cut, on the other hand, has a finite limit $\omega/p \to 0$ (see \eq\nr{eq:llandau}  in Appendix~\ref{app:HTL}) and indeed gives a nonzero, in fact the only nonzero, contribution to $\kappa_\infty(t)$. 

One can see in two ways that the quasiparticle contributions vanish. One way is to simply use the delta function representation~\nr{eq:sincdelta} and then note that the particle contribution to $\dot{\rho}$ in \eq\nr{eq:rhopart} is proportional to $\omega$. Alternatively, if one first integrates over $\omega$ as in \eq\nr{eq:omegaint}, it is the quasiparticle damping term $e^{ -\gamma(p) \Delta t}$ that vanishes in the  $\Delta t \to \infty$ limit. 

Thus, for the heavy quark diffusion coefficient $\kappa_\infty^{\text{SR}}(t)$ using the spectral reconstruction model method, we are left with only the Landau damping contribution, resulting in
\begin{align}
\label{eq:kappaInftyGeneral}
 \kappa_\infty^{\text{SR}}(t)   = \frac{1}{6 N_c} 
 \int \frac{\ud^3p}{\left( 2 \pi \right)^3}\, g^2\langle E E \rangle_{L}(t,t,p)\,\frac{\pi p}{p^2 + m_D^2}.
\end{align}
A straightforward evaluation yields 
\begin{align}
\label{eq:HTLresults}
\kappa_{\infty}^{\text{SR}}\left(Q t = 1500\right) &= 7.4 \times 10^{-5}Q^3\\
\label{eq:HTLresultsb}
 \kappa_{\infty,\mathrm{th. IR}}^{\text{SR}}\left(Q t = 1500\right) &= 6.5 \times 10^{-5} Q^3
 \end{align}
for our measured $\langle E E \rangle_{L}(t,t,p)$ correlator and for the ``thermal IR'' parametrization in \fig\ref{fig:statfun} where the infrared enhancement has been removed, respectively.

To obtain a heavy-quark diffusion coefficient that is accurate to leading logarithmic order, we can use the simpler parametrization 
\begin{equation}
\label{eq:htlEE}
    \langle E E \rangle_{L}^{LL}(t,t,p) = (N_c^2-1)\,T_*\,\frac{ m^2_D}{p^2 + m^2_D} \,\theta(\Lambda-p)\,.
\end{equation}
Then we would get 
\begin{align}
\label{eq:srLL}
 \kappa_{\infty,LL}^{\text{SR}}(t) \;&= \frac{N_c^2-1}{24 \pi N_c}\, m_D^2(t) g^2 T_*(t) \nonumber \\
 & \times \left[\log\left(1+ \frac{\Lambda^2(t) }{m_D^2(t)} \right) - 1 + \frac{m_D^2(t) }{m_D^2(t) + \Lambda^2(t) } \right]  \nonumber \\
 &\approx \frac{N_c^2-1}{12 \pi N_c}\, m_D^2(t) g^2 T_*(t) \log\left( \frac{\Lambda(t) }{m_D(t) } \right),
\end{align}
where in the last line we have left only the leading logarithmic contribution. We leave this equation here as a reference since we will come back to it in the context of the kinetic theory expression discussed in the following.

\subsection{Kinetic theory (KT) framework}
\label{sec:kappaKT}

We can also estimate $\kappa_{\infty}(t)$ in the kinetic theory framework. Here the physical picture is quite intuitive: the heavy quark gains momentum from independent kicks by gluons in the medium. The scattering can be described by a perturbative $Qg\to Qg$ matrix element that, in the limit of large quark mass, is dominated by $t$-channel exchange of a gluon. To arrive at a scattering rate one additionally needs the gluon distribution for the incoming gluons, and the Bose enhancement factor for the outgoing ones (noting that for us $f \gg 1$ so that we can neglect the unity in the Bose enhancement factor $f+1$). We can easily obtain a quantitative expression for the diffusion coefficient following, e.g., the discussion in~\cite{Moore:2004tg} as
\begin{align}
\label{eq:GDMKTKappa}
\kappa_{\infty}^{\text{KT}}(t)&=
\frac{1}{6 M} \int \frac{\mathrm{d}^3 \boldsymbol{k} \mathrm{d}^3 \boldsymbol{k^\prime} \mathrm{d}^3 \boldsymbol{p^\prime}}{\left(2 \pi\right)^9 8 k^0 k^{\prime 0} M} 
\left(2 \pi \right)^3 
\delta^3\left(\bs{p} + \boldsymbol{k^\prime} - \boldsymbol{p^\prime} -\boldsymbol{k} \right) \nn \\
&\times 2 \pi \delta \left(k^\prime - k \right) \boldsymbol{q}^2 \left|\mathcal{M} \right|^2_{\mathrm{gluon}} f(t,k) f(t,k^\prime),
\end{align}
where $M$ is the mass of the heavy quark. The incoming heavy quark momentum is $(M,\bs{p})$ and the outgoing heavy quark momentum is $(M,\boldsymbol{p^\prime})$. The gluon  momenta before and after a collision are given by $(k,\bs{k})$ and $(k^\prime,\bs{k^\prime})$. The transferred momentum is given by $\bs{q} = \bs{p^\prime} - \bs{p}$. 
Compared to the calculation in \cite{Moore:2004tg}, which takes place in a thermal background, we have neglected scatterings from  quarks and replaced the thermal gluon distributions with general ones.

The process is dominated by t-channel gluon exchange. We take the matrix element squared to be
\begin{align}
\left|\mathcal{M} \right|^2_{\mathrm{gluon}} = N_c C_H g^4 16 M^2 k^2 \left( 1+ \cos^2{\left(\theta_{\bs{k}\bs{k}'}\right)}\right) \frac{1}{\left(q^2+m_D^2\right)^2},
\label{eq_matrix_elmt}
\end{align} 
where $C_H = (N_c^2-1)/(2N_c)$ is the color Casimir of the heavy quark and where we introduced an infrared regulator in terms of $m_D^2$. 

The angle between $k$ and $k^\prime$ can be expressed as
\begin{equation}
\cos\left(\theta_{\bs{k}\bs{k}'}\right) = 1 - \frac{\bs{q}^2}{2 \bs{k}^2}.
\end{equation}
Carrying out the integrals that are possible using the delta functions and an angular integral, we can transform this to
\begin{align}
\kappa_{\infty}^{\text{KT}}(t)  &=  \dfrac{ N_c C_H g^4 }{12 \pi^3}  \int_{0}^{\infty} \mathrm{d} k\, k^2   \int_0^{2 k}\mathrm{d}q\, q^3  \left(2-\dfrac{q^2}{k^2} + \dfrac{q^4}{4 k^4}\right) \nonumber \\
& \times   \dfrac{1}{\left(q^2+m_D^2\right)^2}\, f^2\left(t, k \right)
\nonumber \\
&=   \frac{ N_c C_H g^4 }{12 \pi^3}  \int  \mathrm{d} k k^2 
\Big[ 
-\frac{3 m_D^2}{2 k^2}  -\frac{4 k^2}{4k^2+m_D^2} \nonumber \\
& -\frac{m_D^2 \log \left(\frac{m_D^2}{4k^2+m_D^2}\right)}{k^2} 
    +\log \frac{\left(4 k^2+m_D^2\right)}{m_D^2} \nonumber \\ 
& +\frac{3   m_D^4 \log \left( \frac{4 k^2}{m_D^2}+1\right)}{8 k^4}-1 
   \Big]
   f^2\left(t, k \right) . 
   \label{eq:kappa_KT_Tinfty}
\end{align}
This is now an integral that we can carry out numerically. Using the definition of the gluon distribution in \eq\eqref{eq_def_fp}, we get, again at $Qt=1500$, the result
\begin{align}
\label{eq:KTinfiniteTresult}
\kappa_{\infty}^{\text{KT}}\left(t = 1500\right) &= 4.3\times 10^{-5} Q^3 \\
\label{eq:KTinfiniteTresultb}
\kappa_{\infty,\mathrm{th. IR}}^{\text{KT}}\left(t = 1500\right) &= 1.9\times 10^{-5} Q^3\,.
\end{align}
The first one of these is a little bit smaller than, but roughly in line with
the results that we obtain from the SR method, see \eqs\nr{eq:HTLresults} and \nr{eq:HTLresultsb}. The second one is smaller by a factor of more than two.
The effect of removing the infrared enhancement (i.e. the ``thermal IR'' approximation) is larger here than it is in the SR method. This is somewhat paradoxical, since conceptually the kinetic theory calculation is based on the integral over $k$ in \nr{eq:KTinfiniteTresult} being dominated by UV particle like degrees of freedom in $f(k)$, whereas the expression \nr{eq:kappaInftyGeneral} in the SR depends on the longitudinal correlator in the soft momentum region. The effect of the infrared enhancement is large enough that such parametric estimates start becoming unreliable quantitatively, even if they remain true at the leading logarithmic level, as we will discuss next.

Let us now discuss how  the KT and SR approaches are equivalent at the leading logarithmic order for large  $\Lambda/m_D$, again roughly following the discussion in \cite{Moore:2004tg}. We start from the first form in  \eq\eqref{eq:kappa_KT_Tinfty} that has an integral over both $k$ and $q$. Now the momentum of the gluon $k$ is typically of order $\Lambda$, while $q$ is of the order of the Debye scale $m_D$. In the limit $\Lambda \gg m_D$ we can therefore assume that $k \gg q $. This limit enables several modifications that decouple the  $k$ and $q$ integrals. First, at leading order we can replace the upper limit in the $q$ integral by the UV scale $\Lambda$. We can also drop terms $\mathcal{O}\left(q^2 / k^2\right)$ in the integrand. In terms of the original kinetic theory this corresponds to a small-angle approximation where the scattering angle between $k$ and $k^\prime$ is $\cos \theta_{\bs{k}\bs{k}'} \approx 1$. With these approximations we arrive at the same result as the leading logarithmic limit in the SR picture, \eq\nr{eq:srLL}
\begin{align}
 &\kappa_{\infty,LL}^{\text{KT}}(t) \nn \\
 &    = \dfrac{ 2N_c C_H g^4 }{12 \pi^3}  \int_{0}^{\infty} \mathrm{d} k\, k^2   f^2\left(t, k \right) \nonumber   \int_0^{\Lambda}\mathrm{d}q\, q^3  
    \dfrac{1}{\left(q^2+m_D^2\right)^2}
    \nonumber \\
   & =\dfrac{ N_c C_H}{6\pi^3}\,\frac{4\pi^4}{2N_c} \,g^2 T_*\,m_D^2 \int_0^{\Lambda}\frac{\ud q}{(2\pi)^3}\,  
    \dfrac{q}{\left(q^2+m_D^2\right)^2}
    \nonumber \\
    &= \frac{1}{6 N_c} 
 \int \frac{\ud^3q}{\left( 2 \pi \right)^3}\, g^2\langle E E \rangle_{L}^{LL}(t,t,q)\,\frac{\pi q}{q^2 + m_D^2} 
 \nonumber \\
 &= \kappa_{\infty,LL}^{\text{SR}}(t)\,,
 \label{eq:kthtl}
 \end{align}
where we used the integral defining $T_*$ from \eq\nr{eq:effT}
and the assumption that the longitudinal $\langle E E \rangle_{L}^{LL}(t,t,p)$ correlator has the form  $\propto T_*$ that we expect in HTL, \eq\nr{eq:htlEE}.

To summarize, there is a common limit for our SR and KT models. From the KT side, it can be reached by systematically taking the limit of large $\Lambda/m_D$, implying small angle scatterings  and neglecting the exact kinematical limits for the scattering process. From the SR model reaching the common limit requires removing the infrared enhancement that we see in the equal time correlator and replacing it by a functional form that is proportional to $\sim T_*$ which is determined by the hard modes (by the integral~\nr{eq:effT}). For the values of $t$ that we have studied, i.e., the values of the scale separation $\Lambda/m_D$, we see no indication of a disappearance of this infrared enhancement. However, even with the IR enhancement the leading logarithmic limit \eq\nr{eq:srLL} should still be valid, the effect being on the value of the constant under the logarithm. 

Incidentally, the equivalence of the common limit $\kappa_{\infty,LL}(t)$ in the KT and SR frameworks in \eq\nr{eq:kthtl} justifies a posteriori the form of IR regulation we used in \eq\eqref{eq_matrix_elmt}. One arrives at the known leading logarithmic expression of $\kappa_{\infty,LL}^{\text{KT}}(t)$ in thermal equilibrium with temperature $T$ \cite{Laine:2009dd,Moore:2004tg} by setting $T_* \mapsto T$ and $\Lambda \mapsto T$. 

We can use the leading logarithmic limit to estimate the expected scaling behavior for the diffusion coefficient with $t$, which controls the magnitudes of the relevant scales $\Lambda$, $T_*$ and $m_D$ as discussed in Sec.~\ref{sec:cascade}. Based on the scaling behavior of $m_D \sim \Q (\Q t)^{-1/7}$ and $g^2 T_* \sim \Q (\Q t)^{-3/7}$ in the self-similar regime, the scaling behavior of the diffusion coefficient is expected to be 
\begin{equation}
\label{eq:tscaling}
\kappa_{\infty,LL}^{\text{KT}}(t) \approx  \frac{N_c^2-1}{12 \pi N_c}\, m_D^2(t)\, g^2 T_*(t) \log\left[ \frac{\Lambda(t) }{m_D(t) } \right] \sim t^{-5/7},
\end{equation}
up to logarithmic corrections. 

For a generalization to finite time $\kappa^{\text{KT}}(t, \Delta t)$, we allow for a possibility of a nonzero frequency $\omega = k_0^\prime - k_0 = k^\prime - k$, corresponding to finite energy transfer between gluons and the heavy quark. Here one should in principle use the proper HTL expression. We, however, simplify the form by employing the same IR regulator as for $\kappa_{\infty}^{\text{KT}}(t)$ in \eq\eqref{eq_matrix_elmt}. In this case the matrix element becomes 
\begin{align}
 \left|\mathcal{M} \right|^2_{\mathrm{gluon}}(\omega) = N_c C_H g^4\, \frac{4 M^2 (k_0 + k_0')^2 \left( 1 + \cos^2 \theta_{\bs{k}\bs{k}'} \right)}{(q^2-\omega^2+m_D^2)^2},
 \label{matrix_element}
\end{align}
and the frequency dependent expression can be written as 
\begin{multline}
\label{eq:KT_omGen}
\kappa^{\text{KT}}(t, \omega) = \frac{1}{6 M} \int \frac{\mathrm{d}^3 k\, \mathrm{d}^3 k'}{\left(2 \pi\right)^6 8\, k\,k'\,M} 2 \pi \delta(k' - k - \omega)
\\
\times \left(\boldsymbol{k}-\boldsymbol{k}'\right)^2 \left|\mathcal{M} \right|^2_{\mathrm{gluon}}(\omega) f(t,k)f(t,k').
\end{multline}
We carry out this integral by a change of variables from $\bs{q}$ to $\bs{k^\prime}.$ The integrals over the angular variables can be carried out analytically. After Fourier transforming, our expression for the $\Delta t$ dependent diffusion coefficient is 
\begin{align}
\label{eq:KTfiniteTMasterEQ}
&\kappa^{\text{KT}}(t, \Delta t ) = N_c C_H  g^4  \int_0^\infty \mathrm{d} k\, \mathrm{d} k^\prime f(t,k) f(t,k^\prime) (k+k^\prime)^2   \nn \\
 & \times  \Big[m_D^2
   \left(4 k k^\prime+m_D^2\right) \left(-2 m_D^2 (k^2-6 k
   k^\prime+(k^\prime)^2\right)  \nonumber \\
   &-4 k k^\prime \left(k^2-4 k k^\prime+(k^\prime)^2 )+3
   m_D^4\right) \nn  \log \left(\frac{m_D^2}{4 k k^\prime+m_D^2}\right)    \nonumber \\
   & + 4 k   k^\prime \left(-2 m_D^4 (k^2-9 k k^\prime+(k^\prime)^2\right) \nn  \\ 
   & -8 k k^\prime   m_D^2 \left(k^2-4 k k^\prime+(k^\prime)^2\right)  -8 k^2 (k^\prime)^2
   (k-k^\prime)^2+3 m_D^6) \Big] \nn    \\ 
   &\times \Big[384 \pi^4 k^2
   (k^\prime)^2 m_D^2  \left(4 k k^\prime+m_D^2\right) \Big]^{-1}   \frac{ \sin (\Delta t(k^\prime-k))}{(k-k^\prime)}. 
\end{align}
The remaining integrals are evaluated numerically. One recovers $\kappa^{\text{KT}}(t, \Delta t \rightarrow \infty ) = \kappa_\infty^{\text{KT}}(t)$. This procedure has enabled us to get a $\Delta t$-dependent expression from the KT description.

We would only expect this result in \eq\nr{eq:KTfiniteTMasterEQ} to describe the physics at large $\Delta t$. In terms of the different physical ingredients included in the SR description, this finite-$\Delta t$ calculation in kinetic theory still includes only the effect of the longitudinal Landau cut, and even that quite crudely. The gluon exchanged in the kinetic theory calculation is longitudinal and has  a very spacelike four-momentum, because the quark is infinitely heavy and the scattering gluon interacts only with its Coulomb field.
As we will see explicitly in the following, the contribution of the other parts of the spectral function: the quasiparticle poles and the transverse Landau cut, are essential to reproduce the full $\Delta t$ dependence.

The effective Debye mass in the matrix element in Eq.~\eqref{matrix_element}  has been tuned such that it gives the leading-order correct late-time limit but as it has a trivialized 
frequency structure it cannot capture the full leading-order accuracy 
at finite $\Delta t$. 
In order to arrive to full leading order description of the time evolution of $\kappa$ one could further improve the estimate by replacing the simple mass regularization in Eq.~\eqref{matrix_element} by the fully resummed HTL expression (as is done in, e.g., ~\cite{Arnold:2002zm}).

\section{Results}
\label{sec:res}

Here we present our numerical results for the evolution of the heavy-quark diffusion coefficient $\kappa_{\infty}(t)$ and the correlator $\kappa\left(t , \Delta t \right)$ and compare them to results using the SR and KT methods introduced in the previous section. With these tools, we are able to understand distinctive features of these quantities like the origin of the oscillations of $\kappa\left(t , \Delta t \right)$. The comparison to our full numerical data also helps us to assess the quality of the (mainly perturbative) SR and KT methods employed here.

\begin{figure}
\centering
\includegraphics[scale=0.5]{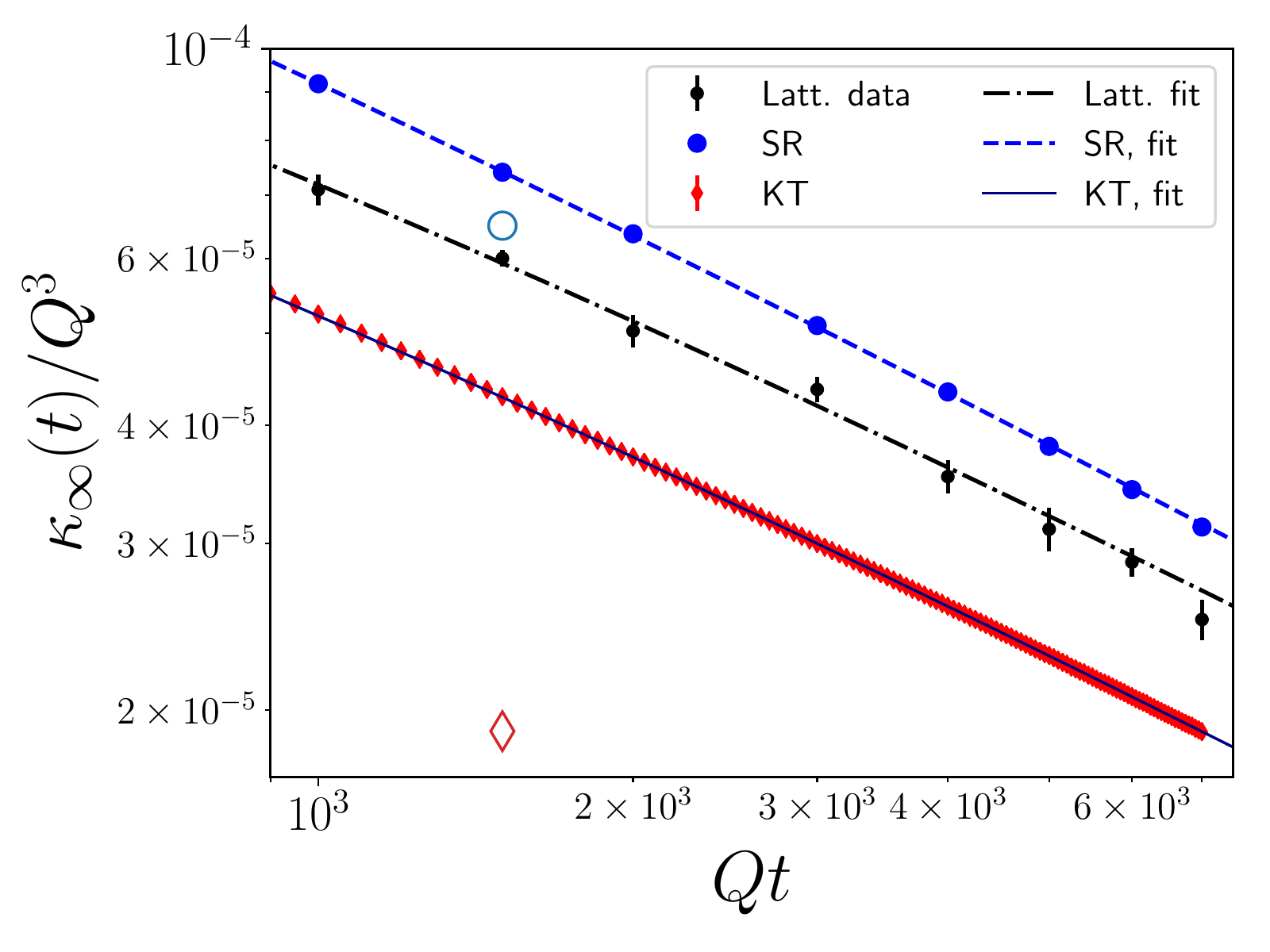}
\caption{Dependence on time of the heavy quark diffusion coefficient $\kappa_{\infty}(t)$. We also show the extracted values from the spectral reconstruction (SR) and kinetic theory (KT) frameworks in the infinite time limit. In addition, we show fits to the lattice data (black dash-dotted line), SR (blue dashed line) and KT (dark-blue line) frameworks whose form is given by \eqref{eq:ourkappa}. The solid points at $Qt=1500$ indicate the numerical values (\eqs \eqref{eq:HTLresults}, \eqref{eq:kappa_fit_values}, and \eqref{eq:KTinfiniteTresult}) discussed in the text. The open points (\eqs \eqref{eq:HTLresultsb} and \eqref{eq:KTinfiniteTresultb}) show the expectations for a thermal infrared spectrum without the enhancement.}
\label{fig:kappa_infty_t}
\end{figure}

\subsection{Time dependence of $\kappa_{\infty}(t)$}
\label{sec:results_t_dep}

We have extracted the values of $\kappa_{\infty}\left(t\right)$ from the numerical calculation using the fitting procedure described above (see \eq\nr{eq_kappa_osc_fit}).%
\footnote{An alternative method to approximate $\kappa_\infty(t)$ is to take an average over $ \Delta t$ in an interval from  $\Delta t_{\text{min}}$ to $\Delta t_{\text{max}}$, with both values $ \gg 1/\Lambda$, as in 
\begin{align}
\label{eq:averaging_procedure}
\kappa_\infty(t) \approx \frac{1}{{\Delta t_{\text{max}}}-{\Delta t_{\text{min}}}}\,\int_{\Delta t_{\text{min}}}^{\Delta t_{\text{max}}} \ud \Delta t\; \kappa(t, \Delta t).
\end{align}
It is easily seen that this is equivalent to cutting off the in principle infinite integration region $\Delta t$ in  \eq\eqref{eq_kappa_infty_def} or \nr{eq:kappa_from_data} by a linear window function going from 1 at $\Delta t_{\text{min}}$ to 0 at $\Delta t_{\text{max}}$. We have observed that the thus extracted values for $\kappa_\infty(t)$ are consistent with our standard extraction method using fits.
}
The resulting $\kappa_{\infty}(t)$ is  shown in \fig\ref{fig:kappa_infty_t} as a function of $t$.

In \fig\ref{fig:kappa_infty_t} we also show the heavy quark diffusion coefficients computed using the SR and KT methods in the \eqs\eqref{eq:kappaInftyGeneral} and \eqref{eq:kappa_KT_Tinfty}, respectively, as discussed in the previous section. 
Both the SR and the KT methods have a time dependence that agrees rather well with our full numerical extraction.
As for the normalization, the SR model slightly overestimates and the KT one underestimates it. Both methods agree with the lattice extraction with roughly 30 \% accuracy.  In the light of the discussion of the common leading logarithmic limit, \eq\nr{eq:kthtl}, we could interpret the lower value for the KT model as being due to importance of the infrared enhancement: the KT model has an implicit assumption that the soft electric field modes have a thermal distribution with a temperature $T_*$ determined by the hard modes as in \eq\nr{eq:effT}, while the SR model actually uses our measured correlator of electric fields at soft momenta, which is larger.

From the leading logarithmic limit we extracted an expectation for the scaling of the diffusion coefficient with time  $t$ in \nr{eq:tscaling}. The resulting power law, $\kappa_{\infty}\left(t\right) \sim t^{-5/7}$, however, receives logarithmic corrections that can be large. 
Also at the earliest values of $t$, it is possible that $\kappa_{\infty}(t)$ suffers from a larger transient effect from the system not yet being fully in the scaling regime. 
Using a simple $t^{-5/7}$ times logarithmic  fit to our data, the evolution of $\kappa_{\infty}(t)$ can be effectively described as
\begin{equation}
    \kappa_{\infty}(t) \approx \frac{\kappa_{\infty}(\Q t=1500)}{C}\left( \frac{\Q t}{1500} \right)^{-5/7}\left( \ln\frac{\Q t}{1500} + C \right)
    \label{eq:ourkappa}
\end{equation}
with two fit parameters
\begin{align}
\label{eq:kappa_fit_values}
    \kappa_{\infty}(Qt=1500) &=  5.90 \times 10^{-5} Q^3 
\nonumber
\\ 
     C &= 4.34   .
\end{align}

It can be instructive to compare our result for  $\kappa_{\infty}(t)$  to the  corresponding thermal value $\kappa_{\infty}^{\text{therm}}$. We emphasize that the classical-statistical framework here cannot be used to follow the system all the way to thermal equilibrium, since one has to remain in the regime of large occupation numbers.  At the times where our system  starts to approach a thermal equilibrium one, it will  simultaneously fall out of the regime of validity of the classical statistical approximation that we are using here. To get a parametrical estimate, this will happen  when  occupation numbers are of order unity $f(t_{\text{therm}},\Lambda) \sim 1$. Due to the scaling behavior $g^2 f(t,\Lambda) \sim (\Q t)^{-4/7}$ this will occur at times that are parametrically $\Q t_{\text{therm}} \sim g^{-7/2}$. Until this time, the non-equilibrium value is larger than the thermal one $\kappa_{\infty}(t) \gg \kappa_{\infty}^{\text{therm}}$ by an inverse power of the coupling $g \ll 1$. 

Let us now try to make this comparison to a corresponding thermal system more quantitative.  The choice to be made in  comparing our result to thermal equilibrium  is to define what is meant by a ``corresponding'' thermal system.  One way to perform such a comparison is to choose a value for the coupling constant $g$ and the energy density $\varepsilon$ and keep them the same for the two systems. For a thermal system these are enough to determine the temperature $T$ and other physical quantities. 
In our case, the combination $g^2\,\varepsilon$ of the coupling and the energy density is fixed by our initial conditions. But subsequently the system also has a time dependence that affects the magnitudes of the different physical scales in the system. Since we cannot follow our simulation to equilibrium we have to choose a meaningful value of $t$ for a comparison to a thermal result. This choice should be made by looking at the value of some other physical quantity.

Since a defining feature of both the thermal system and our overoccupied cascade is the scale separation between a hard scale and a soft Debye scale, a natural way to compare to a thermal system is to compare the Debye scales. At fixed $g^2\,\varepsilon$ we can do this in two ways. One option is to compare our system to a thermal one at the same value of the Debye mass, or equivalently the ratio $m_D^2/\sqrt{\varepsilon}$.
 For a thermal system the energy density and Debye mass are related to the temperature  by $\varepsilon = (N_c^2-1)\pi^2T^4/15$ and $m_D^2 = (N_c/3) g^2 T^2$. We can quantify this Debye scale comparison by defining an ``effective coupling'' with the ratio  $m_D^2/\sqrt{\varepsilon}$ as
\begin{align}
    \tilde{g}^2_{\varepsilon} \,&\equiv 
    \frac{3\pi \sqrt{N_c^2-1}}{N_c \sqrt{15}}\,m_D^2\,\varepsilon^{-1/2} 
    \nonumber \\
    &\approx 2.1\,m_D^2\,\varepsilon^{-1/2} ,
    \label{eq_eff_coupling_eps}
\end{align}
where we have used $N_c = 2$ as in our numerical simulation for the explicit value. 
Another option is to target similar values of the scale separation between the hard and soft scales. This scale separation  can be parametrized by the ratio $m_D/\Lambda$.
For a thermal system with $N_c$ colors  and  no dynamical fermions the energy density, coupling, Debye and hard scales are related by $\varepsilon = (N_c^2-1)\pi^2T^4/15 \approx 1.97\,T^4 \approx 0.031\,\Lambda^4$ and $m_D^2 = (N_c/3) g^2 T^2$. Note that here we define, as in Sec.~\ref{sec:cascade}, the hard scale $\Lambda$ as the momentum scale for which the integrand of the energy density is maximal. For a thermal system this condition ($\ud/\ud p\,(p^3 f_{\mrm{BE}}(p))=0$) leads to $(\Lambda/T)=3(1-\exp(-\Lambda/T))$ and consequently $\Lambda \approx 2.82\,T$. 
We again parametrize  the scale separation $m_D/\Lambda$ in terms of another effective coupling
\begin{align}
    \tilde{g}^2_{\Lambda} \,&\equiv \frac{23.9}{N_c}\,\frac{m_D^2}{\Lambda^2}
    \nonumber  \\ & \approx
    11.9\, \frac{m_D^2}{\Lambda^2} ,
    \label{eq_eff_coupling_lam}
\end{align}
with again $N_c=2$ for the numerical value.
Both of the effective couplings \nr{eq_eff_coupling_eps} and \nr{eq_eff_coupling_lam}  are constructed so that for a thermal system at leading order in $g$ they just give back the gauge coupling:
$\tilde{g}^2_{\varepsilon}=\tilde{g}^2_{\Lambda}=g^2$.

The thermal calculation at leading order (e.g.~\cite{CaronHuot:2007gq}) gives (with $N_c$ colors and no fermions) a  value $\kappa^{\text{therm}} = C_F g^4T^3/(18\pi)N_c (\ln (2T/m_D) - 0.64718)$. In terms of the energy density and the scale separation ratio this gives us 
\begin{align}\label{eq:thkappageps}
\kappa^{\text{therm}} 
&= \frac{N_c^2-1}{36\pi} \left(\frac{15}{(N_c^2-1)\pi^2}\right)^{3/4} \nonumber \\
 & \times \left(\ln \left(\frac{\Lambda}{m_D}\right) - 0.99\right) g^4 \varepsilon^{3/4}
\nn \\ 
&\approx 0.016 \left(\ln \frac{1}{g} + 0.25 \right)g^4 \varepsilon^{3/4}
\end{align}
where in the second equality 
we have taken explicitly $N_c=2$.

For a comparison of our overoccupied system results with this number, we now want to extrapolate our result to a value of $t$ where our measured time dependent Debye scale has a value corresponding to a $\tilde{g}$ that is the same as in a thermal system, i.e., $\tilde{g} \to g$. 
Since our results are usually expressed in terms of $\Q$, we start by relating this quantity to the energy density by $g^2 \varepsilon = 2(N_c^2-1)\int_{\mbf p} p \,g^2 f(t=0,p) \approx 0.0762\,\Q^4$. Thus, we have $\Q^3 \approx 6.9\,(g^2\varepsilon)^{3/4}$ and $\Q = 1.90\,(g^2\varepsilon)^{1/4}$. 
Using the scaling laws in \se\ref{sec:cascade} and the numerical extractions in \eq\eqref{eq_numbers_at_1500}, we get the values for the scale separation  $m_D/\Lambda \approx 0.1\,(\Q t / 1500)^{-2/7}$ and the Debye scale $m_D(t) \approx 0.21\,(\Q t / 1500)^{-1/7}\,\Q \approx 0.40\,(\Q t / 1500)^{-1/7}\,(g^2\varepsilon)^{1/4}.$
Using the definitions of the effective couplings $\tilde{g}_{\varepsilon}$ and $\tilde{g}_{\Lambda}$, \eqs\nr{eq_eff_coupling_eps} and~\nr{eq_eff_coupling_lam}, we can translate values of $Qt$ into values of the effective couplings
\begin{align}
    \left( \frac{\Q t}{1500} \right)^{-1/7} = 1.73\,\frac{\tilde{g}_{\varepsilon}}{g^{1/2}} = 1.703\,\tilde{g}_{\Lambda}^{1/2}\,.
\end{align}
We can then express our result~\nr{eq:ourkappa} for $\kappa_\infty(t)$ in terms of $\tilde{g}_{\varepsilon}$ as 
\begin{equation}\label{eq:noneqkappageps}
        \kappa_\infty(t) \approx 0.0050 \left(\ln \frac{g}{\tilde{g}_{\varepsilon}^2} + 0.148 \right)\tilde{g}_{\varepsilon}^{5}\,g^{-1} \varepsilon^{3/4}.
\end{equation}
or in terms of $\tilde{g}_{\Lambda}$ analogously as
\begin{equation}
\label{eq:noneqkappaglam}
    \kappa_\infty(t) \approx 0.0047 \left(\ln \frac{1}{\tilde{g}_{\Lambda}} + 0.177 \right)\tilde{g}_{\Lambda}^{5/2}\,g^{3/2} \varepsilon^{3/4}.
\end{equation}
The expressions \nr{eq:noneqkappageps} and \nr{eq:noneqkappaglam} involve different powers of the effective couplings due to the fact that the definitions  \nr{eq_eff_coupling_eps} and~\nr{eq_eff_coupling_lam} use $\varepsilon$ and $\Lambda$ that are, in our overoccupied cascade, proportional to $g^{-2}$ and $g^0$.
 It is interesting to see that the coefficients in \nr{eq:noneqkappageps} and \nr{eq:noneqkappaglam} are almost the same. This is a consequence of the values of the Debye and hard scales in \eq\eqref{eq_numbers_at_1500} fortuitously being such that they match the thermal values when extrapolated to late $t$. In other words, it is a coincidence that the conditions $\tilde{g}_{\varepsilon}=g$ and $\tilde{g}_{\Lambda}=g$ are satisfied at not just parametrically, but also numerically at the same values of $t$.

Now, we can make a comparison of the diffusion coefficient  in the overoccupied gluonic cascade to that in a thermal system by comparing \nr{eq:thkappageps} to \nr{eq:noneqkappageps} and \nr{eq:noneqkappaglam}. We firstly re-emphasize that in the overoccupied phase of the evolution, where our classical description is justified, $\tilde{g}_{\varepsilon}\gg g$ and $\tilde{g}_{\Lambda} \gg g$; thus the value of $\kappa_\infty(t) $ is parametrically larger than for a thermal system with the same energy density. We can then extrapolate the time evolution of our system to occupation numbers $f \sim 1$, i.e. the parametric region $\Q t\sim g^{-7/2}$ where the occupation number becomes order one and the classical statistical approximation breaks down.  
This is done in practice by setting  $\tilde{g}_{\varepsilon}\to g$, $\tilde{g}_{\Lambda} \to g$.
With these assignments we observe two features from \eqs \nr{eq:thkappageps}, \nr{eq:noneqkappageps} and \nr{eq:noneqkappaglam}. Firstly, as a consistency check, the powers and logarithms of $g$ are the same in all three cases. Secondly, the value of the coefficient in \nr{eq:noneqkappageps} and \nr{eq:noneqkappaglam} is smaller by a factor of $\sim 3$. This is mostly a result of the IR enhancement and the way we defined the effective couplings using $m_D$. To be more precise: the change in the functional form of $f(p)$ with the IR enhancement increases the value of $m_D$ by 25\% (from a comparison between our result and the ``thermal IR'' parametrization). Since in our classical simulations we have $m_D \sim (Qt)^{\nicefrac{-1}{7}},$ we have to wait for $(Qt)^{\nicefrac{-1}{7}}$ to be 25\% larger to reach the limit $\tilde{g}_{\varepsilon}= \tilde{g}_{\Lambda} = g$, compared to a case without the IR enhancement. During this time,  $\kappa(t) \sim (Qt)^{\nicefrac{-5}{7}}$ becomes smaller by a factor of $1.25^5 \approx 3$. In other words, while the IR enhancement increases the value of $\kappa(t)$, it increases the value of $m_D$ even more, relative to a thermal distribution. This has the effect that when we express the $g$ in  $\kappa \sim g^4 \varepsilon$ in terms of $m_D$, the coefficient is smaller in our case than for a thermal distribution. As long as values are consistently expressed in terms of the appropriate physical scales of the problem, our overall result is thus quite consistent with the perturbative understanding of the microscopic picture of heavy quark diffusion, apart from these effects of the IR enhancement.

\begin{figure}
\centering
\includegraphics[scale=0.5]{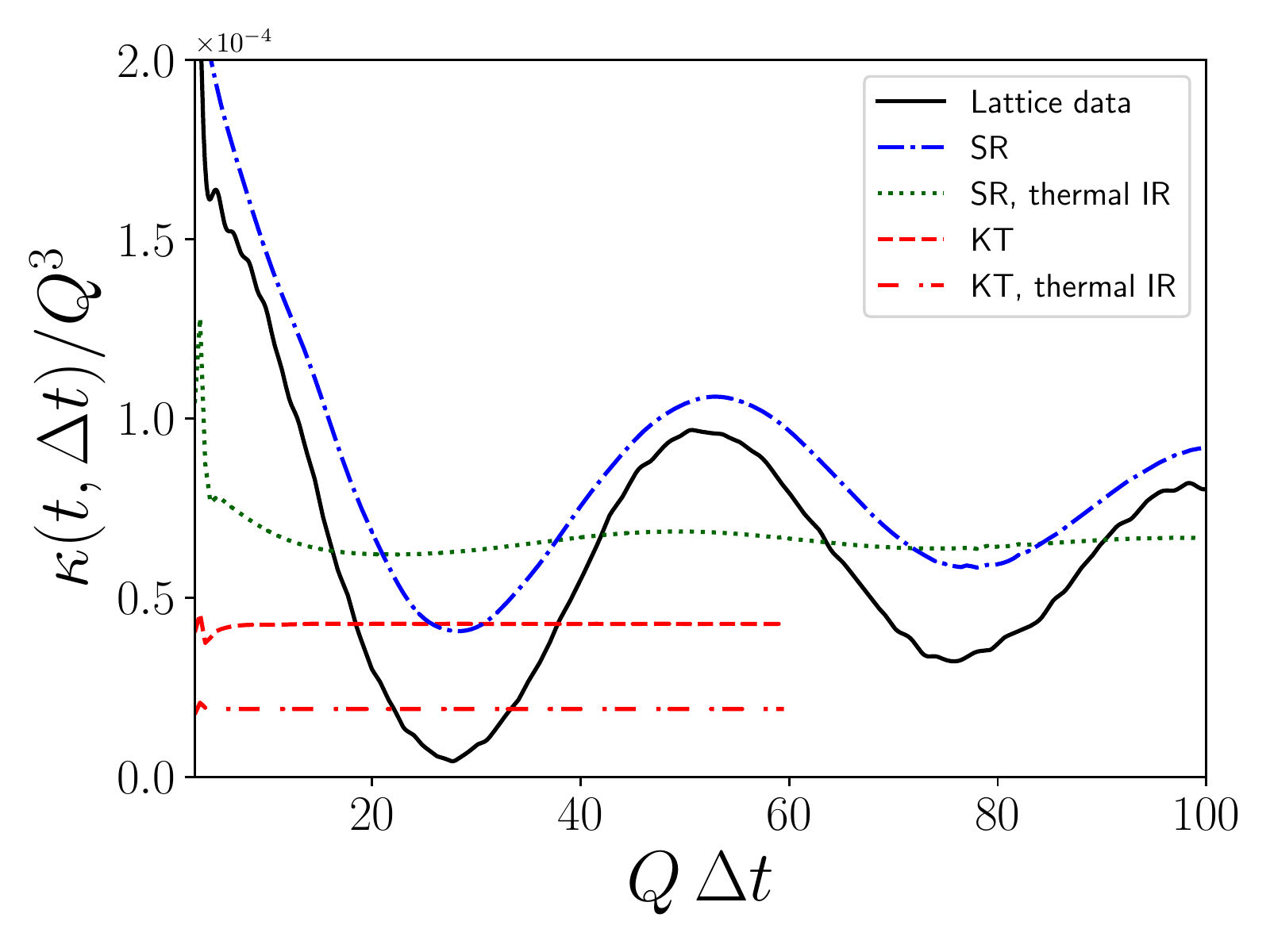}
\caption{Extracted value of $\kappa\left(Q t = 1500 , \Delta t \right)$ as a function of the relative time $\Delta t$. The oscillations in the signal correspond to the plasma frequency, as shown in \fig\ref{fig:osc_freq}. The curve labeled SR is obtained using \eq\nr{eq_kappa_wInt_EE_pInt} with an $\langle EE \rangle $ parametrization  fit to our data in \fig\ref{fig:statfun}, and the one labeled SR thermal IR with the curve extrapolated to a constant $\sim T_*$ in the infrared. The curves labeled KT are obtained using
\eq\nr{eq:KTfiniteTMasterEQ} with the same conventions as for SR. One observes that only the SR curve can reproduce the oscillations reasonably well. As further argued in the text, we take this as a gauge-invariant confirmation of the existence of the IR enhancement.
}
\label{fig:HTL_estimate}
\end{figure}

\begin{figure}
\centering
\includegraphics[scale=0.5]{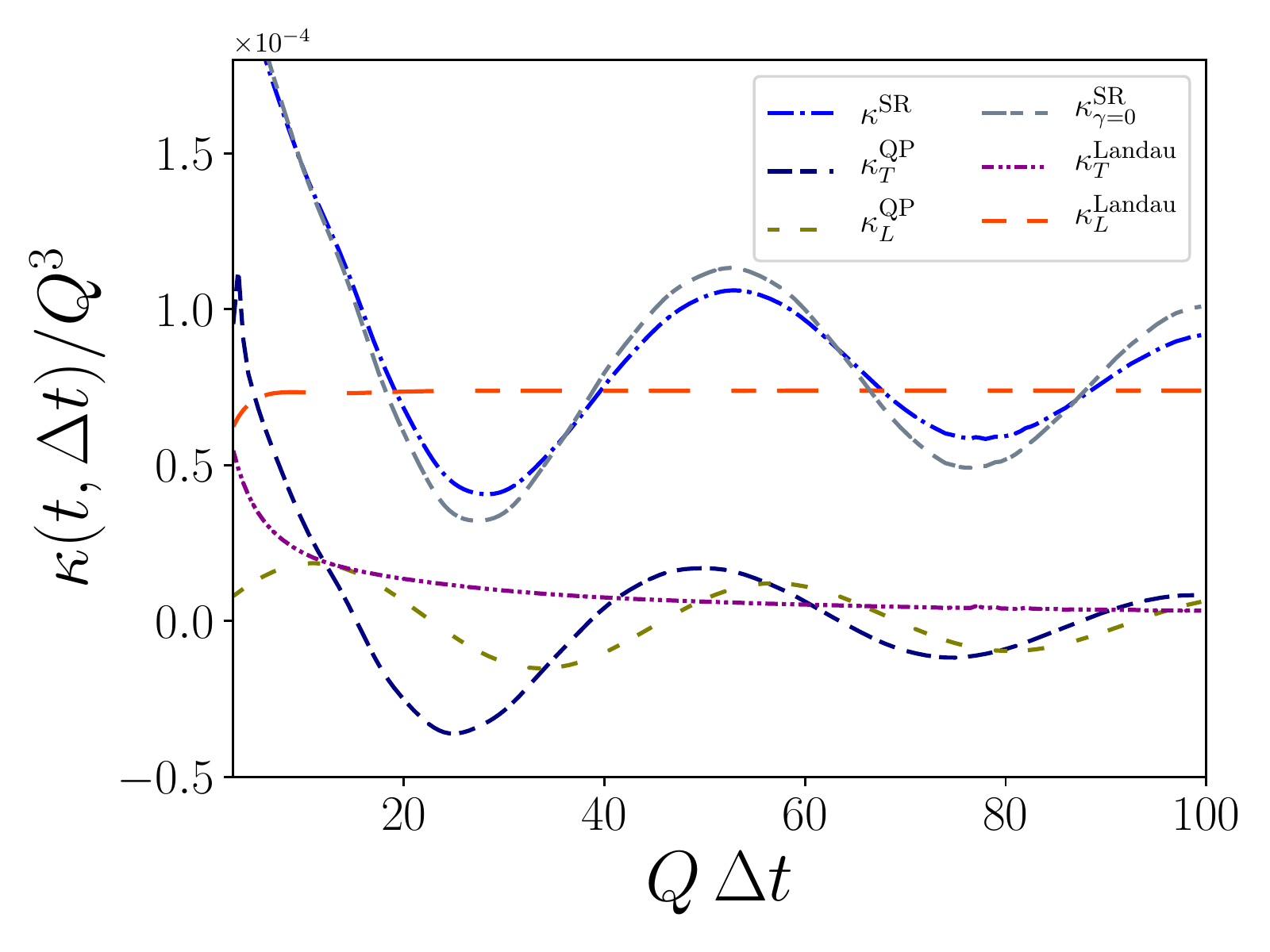}
\caption{Comparison of the different contributions to $\kappa(t, \Delta t)$ computed in the SR framework in \eq\nr{eq_kappa_wInt_EE_pInt}, where the $\langle EE \rangle$ correlator with the extracted IR enhancement is used. The curve for $\kappa^{\text{SR}}$ is also shown in \fig\ref{fig:HTL_estimate}.
We observe that the source of the oscillations at the plasmon mass scale are the quasiparticle contributions, while neglecting the damping rates by setting $\gamma = 0$ has little effect on the oscillations. 
On the other hand, the major contribution to the heavy-quark diffusion coefficient $\kappa_{\infty}$ arises from the longitudinal Landau cut, around which the oscillations proceed. 
}
\label{fig:HTL_break_down}
\end{figure}

\subsection{Understanding the $\Delta t$ dependence of $\kappa\left(t , \Delta t \right)$}
\label{subsec:finiteT}

The self-similar $t$ dependence of the time derivative of $\langle p^2 (t, \Delta t)\rangle$, i.e., of $\kappa\left(t , \Delta t \right)$ has been discussed in \se\ref{sec_kappa_selfsim}. We also showed there that its dependence on $\Delta t$ is well described by a damped oscillating function with oscillation frequency $\wplas$ shifted by $\kappa_{\infty}(t)$, which was parametrized in \eq\eqref{eq_kappa_osc_fit}. Here we study the origin of these oscillations by comparing our data to computations within the SR and KT methods. 

The dependence of $\kappa\left(t , \Delta t \right)$ on $\Delta t$ is shown in \fig\ref{fig:HTL_estimate} for $\Q t = 1500$. 
The black solid curve shows our data while the other curves are obtained using the SR and KT models, \eqs\nr{eq_kappa_wInt_EE_pInt} and~\nr{eq:KTfiniteTMasterEQ}. The blue dashed curve corresponds to the SR model when using the parametrization of the equal time statistical correlation function which agrees with our extracted $\langle EE \rangle(t,t,p)$ correlator in the infrared. The green dotted curve corresponds to the SR calculation in which we use the ``thermal IR'' equal time statistical correlation function, where we have forced the $\langle EE\rangle(t,t,p)$ correlator to a constant $\sim T_*$  in the infrared, see \fig\ref{fig:statfun}.  We observe that the SR method quite well reproduces the oscillations seen in the data, where the shift to slightly larger values of $\kappa$ can be attributed to the overestimation of $\kappa_\infty(t)$ by the SR method, as observed above. However, if the ``IR enhancement'' in the equal time correlation function is removed, the oscillations practically disappear. Likewise, the oscillations seen in the data are not present in the KT model calculation, as could be anticipated based on the discussion in  \se\ref{sec:kappaKT}. 
We take the comparison with the SR model with and without the infrared enhancement in the equal-time correlator as a confirmation of the existence of this infrared enhancement from a manifestly gauge invariant observable. This is one of the main conclusions of our paper.

We can also break down the SR model into different components contributing to $\kappa^{\text{SR}}(t,\Delta t)$, as mentioned in \se\ref{sec:SRmethod}, to get an idea of their relative contributions. This decomposition 
\begin{align}
    \kappa^{\text{SR}} =
    \kappa_{T}^{\mrm{QP}} + \kappa_{L}^{\mrm{QP}} + 
    \kappa_{T}^{\mrm{Landau}} + \kappa_{L}^{\mrm{Landau}}
\end{align}
is shown in \fig \ref{fig:HTL_break_down}, which includes the total $\kappa^{\text{SR}}(t,\Delta t)$ together with the contributions from transverse and longitudinal quasiparticles and the transverse and longitudinal Landau cut. 
We can make two main observations from this figure. 

Firstly, as discussed earlier, the only contribution to the $\Delta t\to \infty$ limit comes from the longitudinal Landau damping which, however, does not show any oscillations (the integration weight being mostly at very small $\omega$ in \eq\eqref{eq_kappa_wInt_EE_pInt}). The transverse Landau cut only contributes at quite small $\Delta t$, and does not exhibit oscillations. 

Secondly, the oscillations are produced by both the transverse and longitudinal quasiparticle contributions, which, however, vanish at $\Delta  t\to \infty$. Their amplitudes decrease due to the specific momentum integration in \eq\eqref{eq_kappa_wInt_EE_pInt} and because of a finite damping rate $\gamma_{T,L}(p)$. The latter seems to have a smaller effect on the evolution. This can be seen in \fig \ref{fig:HTL_break_down} by comparing $\kappa^{\text{SR}}(t,\Delta t)$ to the curve $\kappa^{\text{SR}}_{\gamma = 0}(t,\Delta t)$, where the damping rates have been set to zero in its calculation. Therefore, even LO perturbative calculations, where quasiparticles are given by Delta functions in the spectral function, lead to damped oscillations with frequency $\wplas$ if the  IR enhancement is taken into account.

\section{Conclusions}
\label{sec:conc}

In this paper we have measured the heavy quark diffusion coefficient $\kappa_{\infty}(t)$ in a self-similar overoccupied cascade system, using numerical calculations in classical gluodynamics. In addition to calculating the diffusion coefficient, we have observed strong oscillatory structures in $\Delta t$ in local gauge-invariant electric field correlators $\kappa\left(t , \Delta t \right)$ at the plasmon frequency scale. 

The correlator  $\kappa\left(t , \Delta t \right)$ is, in physical terms, the derivative of the momentum broadening $\langle p^2(t, \Delta t) \rangle$ of a heavy quark traversing the plasma with respect to the time $\Delta t$. In general, we have observed that $\langle p^2(t, \Delta t) \rangle$ rises fast initially on a time scale of $\Delta t \sim 1/\Q$, where $\Q$ is a typical hard scale of the plasma. This initial rise is followed by an approximately linear growth that can be interpreted as due to  momentum  diffusion. The linear growth is modified by additional damped oscillations with the plasmon frequency. While the initial rapid growth is a direct consequence of the decoherence of hard quasiparticles with momenta $\sim \Q$, in order to understand the observed oscillations, dynamics at lower momenta and frequencies have to be taken into account.

Prompted by these numerical observations, we have constructed two models to understand them, which start from a distribution of gluons in the system, obtained numerically from  an equal-time correlator of electric fields. In the first method, that we call spectral reconstruction (SR), we extrapolate our measured equal time correlators from $(\Delta t=0,p)$  to $\omega$-space using a generalized fluctuation-dissipation relation and assuming a HTL structure for the spectral function.  This structure, which we have also observed numerically in \cite{Boguslavski:2018beu}, includes both damped quasiparticle peaks and a low-frequency Landau damping region. Both contributions are needed to explain the full $\Delta t$ dependence of the electric field correlators $\kappa\left(t , \Delta t \right)$. We found that the oscillations at the plasmon frequency are produced by the quasiparticle contributions, whereas the main contribution to the diffusion coefficient arises from the  Landau damping of longitudinal modes in the plasma. In our second method we use a kinetic theory (KT) calculation where the diffusion coefficient is computed utilizing the fact that the process is dominated by $t$-channel gluon exchange. 
We have shown that these two models are equivalent in the leading logarithmic limit. The screened scattering matrix element in the KT model implicitly corresponds to a thermal distribution of longitudinal field correlators in the infrared, with a temperature determined by the hard modes. The SR model, on the other hand, includes an explicit parametrization of these small momentum modes, which enables a better description of the full lattice results.

Using these models we have argued that the oscillations at the plasmon scale can be explained by a larger occupation  of infrared modes $p \lesssim m_D$  in the system than expected from perturbation theory. This feature, that we call IR enhancement, had been observed earlier in the equal-time electric field correlator in Coulomb gauge. We have now demonstrated an independent confirmation for it in a gauge-invariant observable $\ud \langle p^2(t, \Delta t) \rangle / \ud \Delta t$ in a different part of phase space.  These oscillations are also visible in the momentum broadening observable $\langle p^2(t, \Delta t) \rangle$.

The time-dependence of physical quantities in the scaling system tends to follow simple, analytically derivable power laws as a function of time. We have seen that the time dependence of the heavy quark diffusion coefficient is consistent with the $t^{-5/7} \times \log t$-dependence expected from scaling arguments.
Due to the IR enhancement, the contribution of small momentum quasiparticle-like modes below the plasmon frequency scale is larger than one would expect in perturbation theory. Recall that a crucial feature of our (SR) analysis was to assume an HTL form for the spectral function, which is not proportional to the number of quasiparticles in the system, together with a generalized fluctuation-dissipation relation. This means that the IR enhancement leads to an enhancement of the longitudinal statistical correlation function in the Landau damping region $\omega \ll p$ over the perturbative expectation. Consequently the heavy quark diffusion coefficient is larger than one would perturbatively expect based on just the number of hard gluons in the system. In the case of heavy quark diffusion this  enhancement is not a large correction to the the leading logarithmic behavior of the diffusion coefficient,
 but shows up as an oscillatory deviation from the diffusive behavior. In the kinetic theory framework the IR enhancement is a substantial correction. We take this as an indication that kinetic theory becomes less reliable due to the IR enhancement. 

A similar argumentation can also be used to related transport coefficients like the jet quenching parameter $\hat{q}$ and the momentum broadening resulting from it. Therefore, we believe that the effects of overoccupied IR gluonic quasiparticle modes should be taken into account for calculations of other transport coefficients. Similar effects could be important also in  different gluonic systems, including different initial states, anisotropies or in expanding geometry, if the underlying plasma involves infrared enhancement. It would therefore be interesting to study their effect, and thus the effects of such an  infrared enhancement, on other phenomenological observables. 

Our calculation has been done in an extremely weak coupling, classical field limit. To make a connection to phenomenologically relevant values of the parameters, one needs to scale this to realistic values of the coupling, keeping a meaningful set of physical quantities constant. Unsurprisingly, in the strongly overoccupied system the diffusion coefficient is enhanced by inverse powers of the coupling constant compared to a thermal system at the same energy density. 
We have done a more detailed comparison to a thermal system by extrapolating our overoccupied cascade system to a situation where the scale separation between the energy density and the Debye scale are similar.
The results of this exercise, obtainable from a comparison of \eqs\nr{eq:noneqkappageps} and \nr{eq:noneqkappaglam} to \nr{eq:thkappageps}, are intended to be usable by a brave phenomenologist to estimate the effect of an initial overoccupation of gluonic modes in the  pre-equilibrium stage of a heavy ion collision on calculations  of heavy quark diffusion.


\begin{acknowledgments}
  We are grateful to N.\ Brambilla, M.\ A.\ Escobedo, D.\ Müller, S.\ Schlichting and N.\ Tanji for discussions. T.~L.\ is supported by the Academy of Finland, project No. 321840. This work is supported  by the European Research Council, grant ERC-2015-CoG-681707. J.~P. acknowledges the support by the
International Office of the TU Wien and would like to thank Institute
for Theoretical physics for hospitality during part of this work. The content of this article does not reflect the official opinion of the European Union
and responsibility for the information and views expressed therein lies entirely with the authors.  J.~P.\ acknowledges support for travel by the Jenny and Antti Wihuri Foundation. The authors wish to acknowledge CSC – IT Center for Science, Finland, for computational resources.  We acknowledge grants of computer capacity from the Finnish Grid and
   Cloud Infrastructure (persistent identifier
   urn:nbn:fi:research-infras-2016072533 ).

\end{acknowledgments}


\appendix

\section{Lattice checks}
\label{sec_lattice_checks}

\begin{figure}
\centering
\includegraphics[scale=0.5]{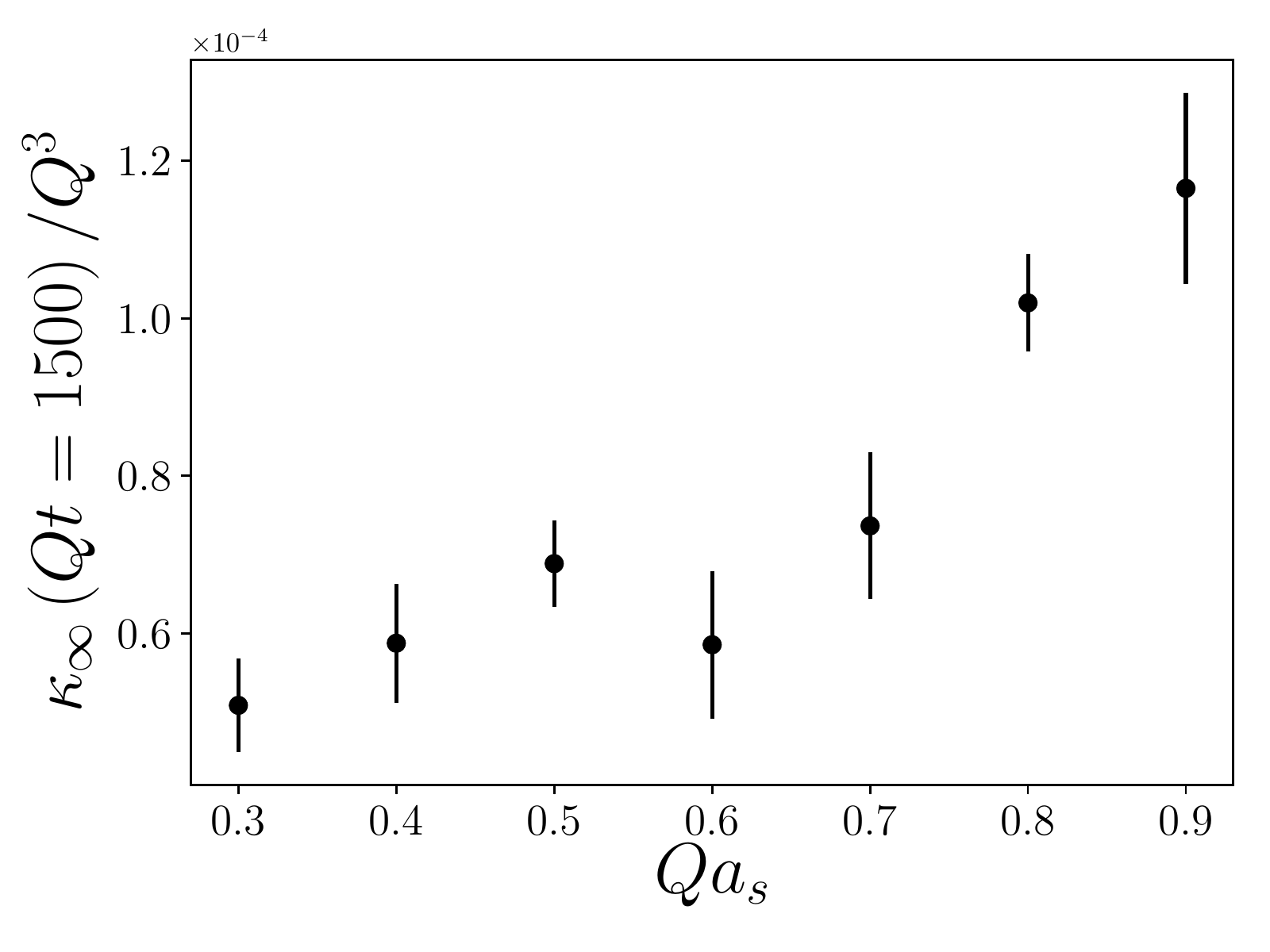}
\caption{Dependence on the lattice spacing of the heavy-quark diffusion coefficient $\kappa_\infty\left(Q t = 1500\right)$, averaged over 4-6 configurations and computed using \eq\nr{eq:averaging_procedure}. Little dependence is visible for lattice spacings $Q a_s \leq 0.6$.}
\label{fig:kappa_vs_Qa}
\end{figure}

The sensitivity to the lattice UV cutoff depends on the lattice spacing $Q a_s$. The dependence of our extraction of the heavy quark diffusion coefficient at $Q t=1500$ on the cutoff is shown in  \fig \ref{fig:kappa_vs_Qa}. The results remain relatively stable up to values $Q a_s = 0.6.$ For larger lattice spacings, we start to observe significant lattice UV cutoff effects, and for $Q a_s=0.8$ the observed value for $\kappa$ has been nearly doubled compared to $Q a_s = 0.3$. The main conclusion in this case is that one has to use lattice spacings $Q a_s \leq 0.6$. This also means that our standard choice $Q a_s=0.5$ is sufficiently small.

\begin{figure}
\centering
\includegraphics[scale=0.5]{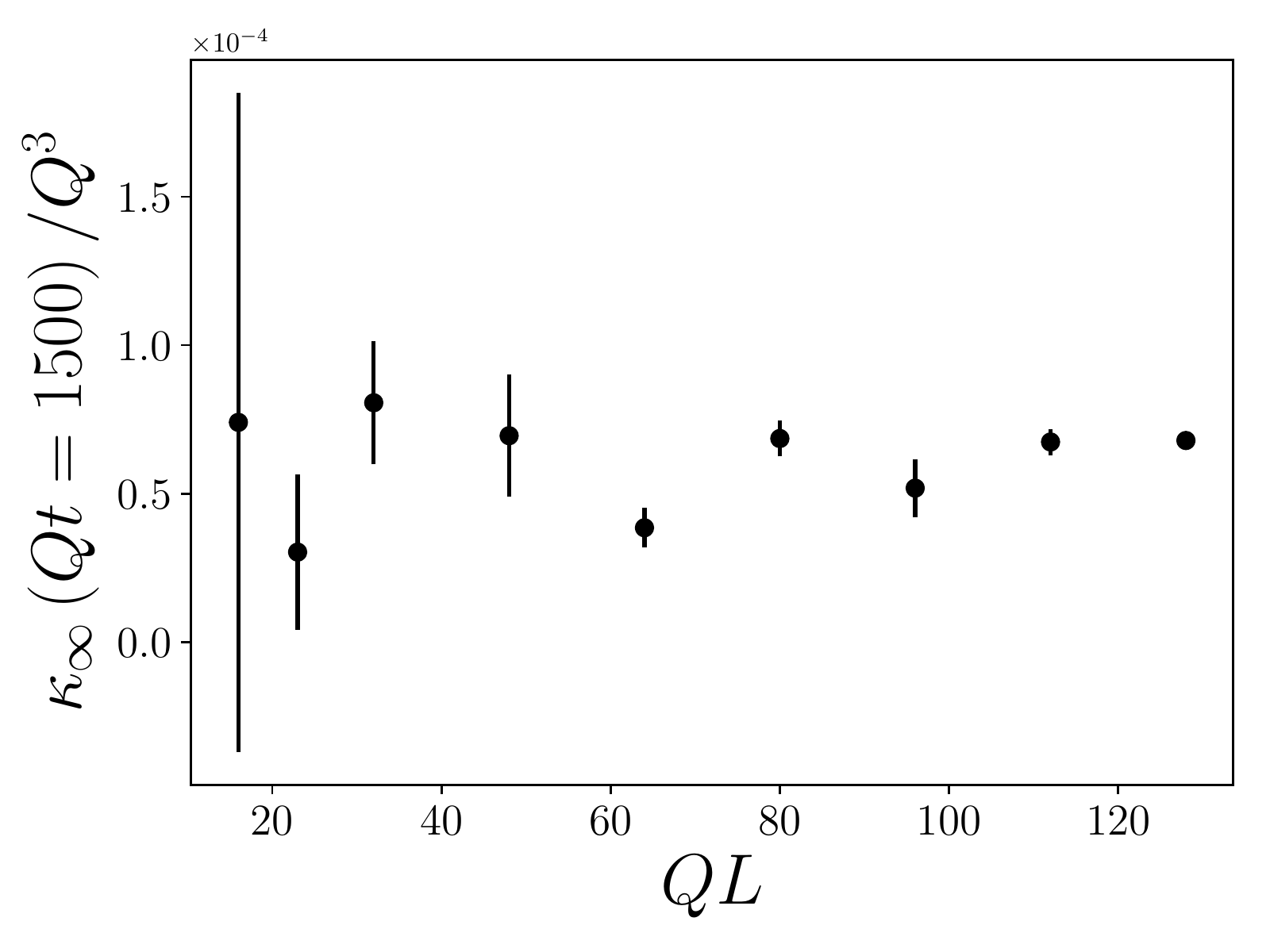}
\caption{Dependence on the lattice length $L \equiv N_s a_s$ of the heavy-quark diffusion coefficient $\kappa_\infty\left(Q t = 1500\right)$, averaged over 4-10 configurations and computed using \eq\nr{eq:averaging_procedure}. For $Q L > 20$, results are seen to be relatively independent of the lattice volume.
}
\label{fig:kappa_vs_QL}
\end{figure}

The infrared cutoff on the lattice is given by the finite size of the system $L \equiv N_s a_s$.
The dependence of the extracted heavy quark diffusion coefficient at $Q t=1500$ on this cutoff  is shown in  \fig \ref{fig:kappa_vs_QL}. The main result is that $\kappa_{\infty}\left(t\right)$ is relatively independent of the IR cutoff, only for lattice sizes $Q L < 20$ we see that our measurements start to break down. Typically our IR cutoff is $\Q L = 130$ ($260^3$ lattice with $Q a_s = 0.5$), which is sufficiently large.

We want our lattice to be large enough that the Debye scale is resolved on the lattice. In our simulations we have $m_D \approx 0.21\,\Q$ at $Q t=1500$. The smallest momentum mode we have available is $k_{\text{min}} = \frac{2 \pi}{L} \approx 0.1\,\Q$ for a $128^3$ lattice. Thus even for $128^3$ lattices we can still accommodate the Debye scale.  Because of the self-similar scaling, we have $m_D \sim t^{\nicefrac{-1}{7}}$ and ultimately at some point the Debye scale will fall below our reach.  Due to the slow power law evolution we expect this to happen roughly at  $Q t = (1500 \times 2^7 ) \approx 190000$, which is much later than our simulation times.

\section{HTL functions}
\label{app:HTL}

In this appendix we will briefly recap the functional forms of the HTL functions which are necessary for the estimation of $\kappa$, in particular for  the decomposition  \eq\nr{eq_rho_contributions}. For more details on these we refer the reader to textbooks on thermal field theory, see e.g., \cite{lebellac_1996} (with a slightly different notation than here). Previously in \cite{Boguslavski:2018beu} we have studied the spectral properties of classical Yang-Mills theory in the self-similar regime, and these functions and their characteristics are discussed there in more detail.

Equal time spectral functions are determined by the sum rules
\begin{align}
\label{eq:spectfun_sumrules}
\dot{\rho}_T^{HTL}\left(t,t , p\right) &= 2 \int_0^\infty \frac{\mathrm{d} \omega}{2 \pi} \dot{\rho}_T^{HTL} \left(t, \omega , p \right) = 1 \\
\dot{\rho}_L^{HTL}\left(t,t , p\right) &= 2 \int_0^\infty \frac{\mathrm{d} \omega}{2 \pi} \dot{\rho}_L^{HTL} \left(t, \omega , p \right) = \frac{ m^2_D(t)}{m^2_D(t) + p^2}. 
\end{align}

The transverse and longitudinal Landau cut contributions are given by the functions
\begin{align}
\label{eq:trlandau}
    &\frac{\dot{\rho}_{T}^{\mrm{Landau}}(t,\omega,p)}{\dot{\rho}_{T}(t,t,p)}  =
    \left(\pi\,p\,\frac{m_D^2}{2}\, x^2(1-x^2)\,\theta\left(1-x^2\right)\right) \nn \\ & \times \Bigg\{
    \Bigg[ p^2 \left(1-x^2\right)   + \frac{m_D^2}{2} \left( x^2+ \frac{x \left(1-x^2\right)}{2} \ln{\left|\frac{x+1}{x-1} \right| }\right)\Bigg]^2 \nn \\ 
    &  +
    \frac{\pi^2}{4} \frac{m_D^2}{4} x^2 \left(1-x^2\right)^2 
\Bigg\}^{-1}
\end{align}    
and
    \begin{align}
    \label{eq:llandau}
    \frac{\dot{\rho}_{L}^{\mrm{Landau}}(t,\omega,p)}{\dot{\rho}_{L}(t,t,p)} & =
    \frac{\pi\,p\,(p^2+ m^2_D)\, \theta\left(1-x^2\right)}{\left[p^2+ m^2_D \left(1- \frac{x}{2} \ln{\left| \frac{x+1}{x-1}\right|} \right) \right]^2 + \pi^2 \frac{m_D^4}{4} x^2}\,,
\end{align}
where $x = \frac{\omega}{p}$. The residues of the quasiparticle peaks are given by the functions
\begin{align}
 Z_T(p) =&\, \frac{\omega_T(p) \left(\omega_T^2(p) - p^2\right)}{m_D^2 \omega_T^2(p) - \left(\omega_T^2(p) - p^2 \right)^2}~, \\
 Z_L(p) =&\, \frac{\left( p^2 + m_D^2 \right)\left(\omega_L^2(p) - p^2 \right)}{\left( p^2 + m_D^2 - \omega_L^2(p)\right)\,\omega_L(p)\,m_D^2}~.
\end{align}

Next we will go through the explicit forms of the dispersion relation $\omega_{T,L}(p)$. In temporal gauge, the retarded transverse and longitudinal propagators are given by
\begin{align}
 \label{eq_GR_HTL}
 G_T(\omega, p) &= \frac{-1}{\omega^2 - p^2 - \Pi_T(\omega/p)} \nonumber \\
 G_L(\omega, p) &= \frac{p^2}{\omega^2}\, \frac{-1}{p^2 - \Pi_L(\omega/p)}\,.
\end{align}
where the transverse and longitudinal gluon self-energy tensors in the HTL framework at LO read 
\begin{align}
 \Pi_T(x) &= m^2\, x\left( x + (1-x^2)Q_0(x) \right) \nonumber \\
 \Pi_L(x) &= -2m^2\, \left(1 - x\,Q_0(x)\right).
\end{align}
Here $Q_0$ is  the Legendre function of the second kind
\begin{align}
 \label{eq_Legendre_fct}
 Q_0(x) = \frac{1}{2}\ln \frac{x+1}{x-1} = \frac{1}{2}\ln \left|\frac{x+1}{x-1}\right| -\frac{i\pi}{2} \theta (1-x^2). 
\end{align}
The dispersion relations of the transverse and longitudinal quasiparticle modes are given by the poles of the retarded propagator. In the general case one has to solve these numerically. It is, however, possible to find approximate solutions for small and large momenta.
For the transverse dispersion relation we have 
\begin{align}
\omega_T &= \sqrt{\wplas^2 + \nicefrac{6}{5}\;p^2}, \quad p \ll m_D\\
\omega_T &= \sqrt{m^2+ p^2}, \quad p \gg m_D,
\label{eq:omegaT_approximations}
\end{align}
while for the longitudinal dispersion relation the corresponding expressions are
\begin{align}
\omega_L &= \sqrt{\wplas^2 + \nicefrac{3}{5}\; p^2}, \quad p \ll m_D \\
\omega_L &= p\left(1 + 2 \exp\left( -\frac{p^2 + m^2}{m^2}\right)\right), \quad p \gg m_D.
\label{eq:omegaL_approximations}
\end{align}

For the damping rates $\gamma_{T,L}(p)$, we parametrize the data we have previously extracted (see \cite{Boguslavski:2018beu}, \fig 9). Since we could not observe any differences between transverse and longitudinal damping, we use the same parametrization for both polarizations. In perturbation theory the damping rate is a next to leading order effect. Perturbatively it has been computed at $p=0$ \cite{Braaten:1990it} and also for momenta of the order of the hard scale \cite{Pisarski:1993rf}.

\bibliographystyle{JHEP-2modlong}
\bibliography{spires}
\end{document}